\documentstyle[a4,amstex]{article}
\newcommand{\real}{{\bold R}}
\newcommand{\cplx}{{\bold C}}
\newcommand{\zint}{{\bold Z}}
\newcommand{\nat}{{\bold N}}

\newcommand{\cplxN}{\cplx^N}
\newcommand{\F}{F_{N,n}}

\newcommand{\cz}{\cplx[z_1^{\pm 1},\dots,z_n^{\pm 1}]}
\newcommand{\cx}{\cplx[x_1^{\pm 1},\dots,x_n^{\pm 1}]}
\newcommand{\gl}{ {{\frak g} {\frak l}}}
\newcommand{\glN}{ {\frak g}{\frak l}_N}

\newcommand{\id}{ {\mathrm i}{\mathrm d}}
\newcommand{\vac}{{\mathrm v}{\mathrm a}{\mathrm c}} 
\newcommand{\odd}{{\mathrm o}{\mathrm d}{\mathrm d}}
\newcommand{\even}{{\mathrm e}{\mathrm v}{\mathrm e}{\mathrm n}}
\newcommand{\exxp}{{\mathrm e}}
\newcommand{\p}{\partial}
\newcommand{\sprod}[2]{( \; #1 \;, \;#2 \; )}
\newcommand{\sprodd}[2]{\langle \; #1 \;, \;#2 \;   \rangle}
\newcommand{\sproddd}[2]{\langle\langle \; #1 \;, \;#2 \;  \rangle\rangle}

\newcommand{\s}{\sigma}

\renewcommand{\o}{\omega}
\newcommand{\sgn}{{\mathrm s}{\mathrm g}{\mathrm n}}
\newcommand{\ov}[1]{\overline{#1}}
\newcommand{\un}[1]{\underline{#1}}

\newcommand{\LC}{{\cal L}}

\newcommand{\setN}{\{1,\dots,N\}}

\newcommand{\halmos}{\rule{5pt}{5pt}}
\newcommand{\ba}{\begin{array}}
\newcommand{\ea}{\end{array}}
\newcommand{\bqq}{\begin{equation}}
\newcommand{\eqq}{\end{equation}}
\newcommand{\bqa}{\begin{eqnarray}}
\newcommand{\eqa}{\end{eqnarray}}
\newcommand{\bqas}{\begin{eqnarray*}}
\newcommand{\eqas}{\end{eqnarray*}}

\numberwithin{equation}{section}

\newenvironment{pf}{\noindent {\it {\normalsize P}\normalsize{roof.}}  
\normalsize\hskip 5pt}{\hfill\halmos}

\begin{document}

\begin{titlepage}
\pagestyle{empty}
\flushright{ }
\begin{center}
\mbox{} \\
\vspace{2.7cm}
\begin{Large}
{\bf Yangian Gelfand-Zetlin Bases, $\glN$-Jack Polynomials \\   
and computation of Dynamical Correlation Functions  in the \\ 
Spin Calogero-Sutherland Model }
\end{Large} \\
\vspace{1.5cm}
\large{ Denis Uglov}   
\\  Research Institute for Mathematical Sciences\\
Kyoto University, Kyoto 606, Japan  \\ 
e-mail: duglov@@kurims.kyoto-u.ac.jp  \\ 
\vspace{1cm}
February  1997  \\

\vspace{2cm}
\begin{abstract}
We consider the $\glN$-invariant Calogero-Sutherland Models with $N=1,2,3,\dots\;$ in a unified  framework, which is the framework of Symmetric Polynomials. By the framework we mean an isomorphism between the space of states of the  $\glN$-invariant Calogero-Sutherland Model and the space of Symmetric  Laurent  Polynomials. In this framework it becomes apparent that all the $\glN$-invariant Calogero-Sutherland Models are manifestations of the same entity, which is the commuting family of Macdonald Operators. Macdonald Operators depend on two parameters $q$ and $t$. The Hamiltonian of $\glN$-invariant Calogero-Sutherland Model belongs to a degeneration of this family in the limit when both $q$ and $t$ approach the $N$th elementary root of unity. This is a generalization of the well-known situation in the case of Scalar  Calogero-Sutherland Model $(N=1).$ 

In the limit the commuting family of Macdonald Operators is identified with the maximal commutative sub-algebra in the Yangian action on the space of states of the  $\glN$-invariant Calogero-Sutherland Model. The limits of Macdonald Polynomials which we call $\glN$-Jack Polynomials are eigenvectors of this sub-algebra and form  Yangian Gelfand-Zetlin bases in irreducible components of the Yangian action. The $\glN$-Jack Polynomials describe the orthogonal eigenbasis of  $\glN$-invariant Calogero-Sutherland Model in exactly the same way as Jack Polynomials describe the orthogonal eigenbasis of the Scalar Model $(N=1).$ For each known property of Macdonald Polynomials there is a corresponding property of $\glN$-Jack Polynomials. As a simplest application of these properties we compute two-point Dynamical Spin-Density and Density Correlation Functions in the $\gl_2$-invariant Calogero-Sutherland Model at integer values of the coupling constant. 
\end{abstract}

\end{center}
\end{titlepage}

\section{Introduction}
In this paper we study the spin generalization of the Calogero-Sutherland Model \cite{Sutherland} which was proposed in \cite{BGHP}. This Model describes $n$ quantum particles with coordinates $ y_1,\dots,y_n $ moving along a circle of length $L$ $( 0 \leq y_i \leq L)$. Each particle carries a spin with $N$ possible values, and the dynamics of the Model are governed by the Hamiltonian   
\begin{equation}
\ov{H}_{\beta,N} = -\frac{1}{2}\sum_{i=1}^n \frac{\p^2}{\p y_i^2} + \frac{\pi^2}{2L^2}\sum_{1\leq i\neq j\leq n}\frac{\beta(\beta + P_{ij})}{\sin^2\frac{\pi}{L}(y_i - y_j)} \label{eq:hamiltonian}
\end{equation}
where integer $\beta > 0$ is a coupling constant and the $P_{ij}$ is the spin exchange operator for particles $i$ and $j.$ As pointed out in \cite{BGHP} it is convenient to make a gauge transformation of (\ref{eq:hamiltonian}) by taking   $W = \prod_{1\leq i<j\leq n}\sin\frac{\pi}{L}(y_i - y_j)$ and defining the gauge-transformed Hamiltonian $H_{\beta,N}$ by 
\begin{equation}
H_{\beta,N} = \frac{L^2}{2\pi^2} W^{-\beta} \ov{H}_{\beta,N} W^{\beta}.  \label{eq:H0}
\end{equation}
If we  set $z_j = \exp(\frac{2\pi i}{L}y_j),$ then $H_{\beta,N}$ acts on the vector space     
\begin{equation}
\F := \left( \cz \otimes (\otimes^n \cplxN ) \right)_{antisymm}
\end{equation}
where $antisymm$ means total antisymmetrization. In this paper we always will be working with the gauge-transformed Model which has $\F$ as its space of quantum states. This space of states is a Hilbert space with a $\beta$-dependent scalar product ${\sprod{\cdot}{\cdot}}_{\beta,N}$ which we describe in section \ref{sec:scalar}.       

It is well-known that the scalar version of the Model $(N=1)$ is best understood in the framework of symmetric polynomials. In this case the space of states $F_{1,n}$ is naturally isomorphic -- by multiplication with the Vandermonde determinant --  to the vector space of symmetric Laurent polynomials, and the orthogonal eigenbasis of $H_{\beta,1}$ is described by Jack Polynomials with parameter $ \frac{1}{\beta + 1} $ in notations of Macdonald's book \cite{MacBook}. Properties of Jack Polynomials known in mathematical literature (e.g. \cite{MacBook,Stanley}) are of paramount importance for the scalar Calogero-Sutherland Model, for a rather straightforward  application of these properties  allows to compute two-point Dynamical Correlation Functions \cite{Ha,LPS,Konno}.      

The main observation of the present paper is that, from our viewpoint,  the Spin Model with arbitrary $N$ is best understood  in the framework of {\em symmetric polynomials} as well. At the first glance the Model with $N \geq 2$ seems to have little to do with symmetric polynomials since it has two disparate types of degrees of freedom: coordinates and spins. It turns out, however, that there is an isomorphism between the Hilbert space $\F$ and the space of symmetric Laurent polynomials such that an orthogonal eigenbasis of  $H_{\beta,N}$ is described by symmetric polynomials which are natural generalizations of Jack Polynomials, and which  we call $\glN$-{\em Jack Polynomials}. 

These symmetric polynomials are generalizations of Jack Polynomials in the sense that they are also limiting cases of Macdonald Polynomials \cite{MacBook}. The Jack Polynomial is a limit of the Macdonald Polynomial  when both parameters in the latter approach 1 \cite{MacBook}. And the  $\glN$-Jack Polynomial is a limit of the Macdonald Polynomial when both parameters in the latter approach the $N$th root of unity $\o_N = \exp(\frac{2\pi i}{N}).$ 

It is clear that for the   $\glN$-Jack Polynomials one can derive analogues of all properties known for Macdonald polynomials just by taking the limit. In precisely the same way as in the case $N=1$ these properties are straightforward to apply in order to compute two-point Dynamical Correlation Functions in the Spin Calogero-Sutherland Model with arbitrary $N.$ As an example, in this paper we give a derivation of Spin-Density and Density two-point Dynamical Correlation Functions for the case of $N=2$ and non-negative integer $\beta.$          

Let us now outline our approach and results in a more detailed manner.  

\subsection{Main result}

As we have already discussed, we treat the Spin Calogero-Sutherland Model with non-negative integer $\beta$ by mapping it isomorphically onto a problem formulated in the language of symmetric polynomials. 

To be more precise, let $\Lambda_n^{\pm} = ( \cplx[x_1^{\pm 1},\dots, x_n^{\pm 1}])^{S_n}$ be the vector space of symmetric Laurent polynomials in variables $x_1,\dots,x_n$ with complex coefficients. For any Laurent polynomial $f$ $=$ $f(x_1,\dots,x_n)$ we will denote by $[f]_1$ the constant term in $f.$ And we will use the bar to denote the complex conjugation. The vector space $\Lambda_n^{\pm}$ is equipped with a scalar product which we denote by ${\sprodd{\cdot}{\cdot}}_{\beta,N}.$ This scalar product is defined in terms of the weight function 
\begin{equation}
\Delta(\beta,N; x_1,\dots,x_n ) :=  \prod_{1\leq i \neq j \leq n}( 1 - x_i^{N}x_j^{-N})^{\beta}
( 1 - x_ix_j^{-1})
\end{equation}
and its value on any two symmetric Laurent polynomials $f$ and $g$ is given by the formula   
\begin{equation}
{\sprodd{f}{g}}_{\beta,N} = \frac{1}{n!} \left[ \overline{f(x_1^{-1},\dots,x_n^{-1})}
\Delta(\beta,N; x_1,\dots,x_n )g(x_1,\dots,x_n)\right]_1 .
\end{equation}
Thus the Hilbert space structure on $\Lambda_n^{\pm}$ is defined. 

On the other hand the space of states $\F$ is also a Hilbert space with the scalar product ${\sprod{\cdot}{\cdot}}_{\beta,N}$ which we  define in section \ref{sec:scalar}.  

The Hilbert space $\Lambda_n^{\pm}$ has an orthogonal basis whose elements are parameterized by an integer number $r$ and a partition $\lambda$ with length less or equal to $n-1.$ The elements of this basis are  
\begin{equation}
 (x_1 \cdots x_n)^r P_{\lambda}^{(N\beta + 1,N)}(x_1,\dots,x_n)
\end{equation}
where for any positive real number $\gamma$ the symmetric {\em polynomial} $P_{\lambda}^{(\gamma,N)}(x_1,\dots,x_n)$ is defined as the following limit of the Macdonald Polynomial $P_{\lambda}(q,t; x_1,\dots,x_n )$ (Cf. \cite{MacBook}):
\begin{equation}
P_{\lambda}^{(\gamma,N)}(x_1,\dots,x_n) = \lim_{p \rightarrow 1} P_{\lambda}(\o_N p, \o_N p^{\gamma} ; x_1,\dots,x_n ).
\end{equation}
Here $\o_N$ is the $N$th elementary root of unity. We will call the polynomial $P_{\lambda}^{(\gamma,N)}$ $=$ $P_{\lambda}^{(\gamma,N)}(x_1,\dots,x_n)$ a $\glN$-{\em Jack Polynomial} since when $N=1$ this polynomial is nothing but the Jack Polynomial $P_{\lambda}^{(\gamma^{-1})}$ in notations of Macdonald \cite{MacBook}. Here it may be useful to observe that with $\gamma = N\beta + 1$ the scalar product ${\sprodd{\cdot}{\cdot}}_{\beta,N}$ is just the limit of the scalar product for Macdonald Polynomials \cite{MacBook}. 

Now we formulate our main result. \\ \mbox{} \\ \mbox{} 

\noindent {\bf {\Large Main result }}

\begin{em}
\noindent In the Hilbert space of states $\F$ of the Spin Calogero-Sutherland Model with $\beta > 0$ there exists an  orthogonal eigenbasis of the Hamiltonian $H_{\beta,N}.$ The elements of this eigenbasis $X_{r,\lambda}^{(\beta,N)}$ are parameterized by an integer $r$ and a partition $\lambda$ with length less or equal to $n-1.$  

Moreover for integer non-negative $\beta$ there exists an isomorphism $\Omega$ of Hilbert spaces $\F$ and $\Lambda_n^{\pm}$ such that  
\begin{equation}
\Omega(X_{r,\lambda}^{(\beta,N)}) = (x_1 \cdots x_n)^r P_{\lambda}^{(N\beta + 1,N)}(x_1,\dots,x_n)
\end{equation}
where $P_{\lambda}^{(N\beta + 1,N)}(x_1,\dots,x_n)$ is the $\glN$-Jack Polynomial.
\end{em} \\
\mbox{} 

Let us make two comments. When $N=1$ the statement above is well-known. In this case acting with the isomorphism $\Omega$ amounts simply to dividing a skew-symmetric Laurent polynomial by the Vandermonde determinant so that the result is a symmetric Laurent polynomial. 

Another comment is about severity of the restriction that $\beta$ be an integer. We need this restriction in order to be able to carry out all our proofs in a completely algebraic manner -- so as not to deal with questions of convergence of various integrals. We conjecture that all our results and formulas, in particular the result above, are valid for all real positive $\beta$ as well, modulo  some evident modifications. We do not however have necessary proofs which  would require analytical considerations.    

\subsection{Yangian Gelfand-Zetlin bases in the Spin Calogero-Sutherland Model}
Let us now explain how we specify the eigenbasis $\{ X_{r,\lambda}^{(\beta,N)} \}$ in the previous section. 

It is known from the work \cite{BGHP} that the space of states $\F$ admits an action of the algebra $Y(\glN)$ -- the  Yangian of $\glN$ \cite{Drinfeld,NT} such that this action commutes with the Hamiltonian $H_{\beta,N}.$ The Yangian $Y(\glN)$ has a maximal commutative sub-algebra $A(\glN)$ \cite{Cherednik,NT} which is generated by centers of all sub-algebras in the chain   
\begin{equation}
 Y(\gl_1) \subset Y(\gl_2) \subset \dots \subset Y(\gl_N), 
\end{equation}
where Yangians $ Y(\gl_1) \subset Y(\gl_2) \subset \dots \subset Y(\gl_{N-1})$ are realized inside $ Y(\gl_N)$ by the standard embeddings \cite{NT} (see section \ref{sec:yangian}). 

The Hamiltonian  $H_{\beta,N}$ is known \cite{BGHP} to belong to the center of the $Y(\glN)$-action on $\F$ which is generated by the Quantum Determinant. This means that the Hamiltonian belongs to the commutative family of operators $A(\glN;\beta)$ which give action of the sub-algebra $A(\glN)$ on $\F.$ Because of this it is natural to concentrate our attention on this commutative family rather than on the Hamiltonian alone. Doing this has two advantages. 

The first is that the spectrum of the commutative family $A(\glN;\beta)$ is {\em simple} \cite{TU} unlike the spectrum of the Hamiltonian which has degeneracy coming from the Yangian symmetry. Therefore an eigenbasis of  $A(\glN;\beta)$ is defined uniquely up to normalization of eigenvectors. It is precisely the eigenbasis  $\{ X_{r,\lambda}^{(\beta,N)}\}.$   

The second advantage is that since each operator from the family $A(\glN;\beta)$ is self-adjoint relative to the scalar product ${\sprod{\cdot}{\cdot}}_{\beta,N}$ \cite{TU}, the elements of the eigenbasis  $\{ X_{r,\lambda}^{(\beta,N)}\}$ are mutually orthogonal automatically because of the simplicity of the spectrum.   

According to \cite{NT} Yangian representations where action of the sub-algebra $A(\glN)$ is diagonalizable are called {\em tame}, and  $A(\glN)$-eigenbases in irreducible tame representations are called Yangian Gelfand-Zetlin bases. As was established in \cite{TU} (see also sections \ref{sec:decomposition} and \ref{sec:a-eigenbasis} of the present paper)  the space of states $\F$ is tame and completely reducible Yangian representation. The basis $\{ X_{r,\lambda}^{(\beta,N)}\}$ is just a direct sum of Yangian Gelfand-Zetlin bases of the irreducible components of $\F.$    

From the main result given in section 1.1 and the above discussion it is apparent that the image of the commutative family  $A(\glN;\beta)$ under the isomorphism $\Omega$ is nothing but the degeneration of the commutative family of Macdonald Operators of which Macdonald Polynomials are eigenvectors \cite{MacBook}.

\subsection{Computation of the Spin-Density and Density Dynamical Correlation Functions for $N=2$ }
One of the consequences of our main result is that we may compute Spin-Density and Density Dynamical  Correlation Functions in the Spin Calogero-Sutherland Model $(N \geq 2)$ in a practically the same way as in the Scalar Calogero-Sutherland Model $(N = 1)$ \cite{Ha,LPS,Poly}. The only extra work which has to be done in the case $N \geq 2$ is to identify the ground state, and identify the operators on the space of symmetric Laurent polynomials that are obtained when we twist the Spin-Density and Density operators with the isomorphism $\Omega.$ 

The identification of the ground state is a painstaking process of finding  the state with the lowest energy eigenvalue which, however, is easy to do when $N=2$ and the number of particles in the Model $n$ is even  such that $n/2$ is odd. In this case we find that the ground state is a basis in a one-dimensional Yangian representation and is, in particular, a spin singlet. It is mapped by the $\Omega$ into the Laurent polynomial 1. 

Images of  the   Spin-Density and Density operators under the isomorphism $\Omega$ are just  power-sums acting as multiplication operators on the space of symmetric Laurent polynomials. In particular, when $N=2$ the  Spin-Density is mapped into a sum of odd power sums, and the Density is mapped into   a sum of even power sums.   

This being established, our computation of the Correlation Functions for $N=2$ follows exactly the computation for the scalar case \cite{Ha,LPS} with the only difference that we use the $\gl_2$-Jack Polynomials instead of Jack Polynomials. As in  the scalar case our result for e.g. Spin-Density  Correlation Function is  represented as a sum over all partitions $\lambda$ of length less or equal to $n$ such that for non-negative integer $\beta$ the summand vanishes if the diagram of $\lambda$ contains the square with leg-colength 1 and arm-colength $2\beta + 1$ \cite{MacBook}.     

In the present paper we do not consider the thermodynamic limit of these Correlation Functions.

Another problem which we do not consider in this paper is a computation of Green's  functions \cite{Ha,LPS}. This problem, however, also appears to be tractable in our approach if we take into account the Cauchy formula for $\glN$-Jack Polynomials:
\begin{equation}
\sum_{\lambda} P_{\lambda}^{(\gamma,N)}(x_1,\dots,x_n)P_{\lambda'}^{(\gamma^{-1},N)}(y_1,\dots,y_n) = \prod_{i,j =1}^n ( 1 + x_i y_j), 
\end{equation}
where $\lambda'$ is the partition conjugated to $\lambda.$ This Cauchy formula is obviously the limit of the corresponding formula for Macdonald Polynomials \cite{MacBook}.

\subsection{Plan of the paper}
Let us now outline the plan of the present paper. In section \ref{sec:preliminary} we summarize some notational conventions to be used troughout the paper. We also define here the {\em wedge } vectors, or simply, wedges which form a basis of the space of states $\F.$ The isomorphism $\Omega$ mentioned in section 1.1  will map these wedges into Schur polynomials \cite{MacBook}, or, more precisely, into  natural extensions of Schur polynomials which form a basis in the space of symmetric Laurent polynomials.      

In section \ref{sec:scalar} we define an appropriate scalar product on the space $\F.$ 

In section \ref{sec:gauge} we describe the gauge-transformed Hamiltonian $H_{\beta,N}$ and recall its relation to the Dunkl operators following \cite{BGHP}.

In section \ref{sec:eigenbasis1} we construct a certain eiegenbasis of $H_{\beta,N}.$ This eigenbasis is not orthogonal, but it plays an important role in subsequent considerations.   

Sections \ref{sec:yangian} - \ref{sec:a-eigenbasis} deal with the Yangian symmetry of the Spin Calogero-Sutherland Model and the Yangian Gelfand-Zetlin bases.
In section \ref{sec:yangian} we summarize properties of the Yangian algebra giving particular attention to the maximal commutative subalgebra $A(\glN).$

In section \ref{sec:yangian in scsm} we recall the definition of the Yangian action in  the Spin Calogero-Sutherland Model following \cite{BGHP}.

Section  \ref{sec:decomposition} is the summary of the results of the paper \cite{TU} concerning the decomposition of the space of states $\F$ into irreducible Yangian subrepresentations. 

In section \ref{sec:a-eigenbasis} we describe the eigenbasis of the commutative family $A(\glN;\beta)$ in the space of states. This is the basis $\{ X_{r,\lambda}^{(\beta,N)} \}$ mentioned in the section 1.1 of this introduction.

Sections \ref{sec:notations} - \ref{sec:identification} are concerned with formulation of the Spin Calogero-Sutherland Model in the language of symmetric polynomials. In section \ref{sec:notations} we summarize notations concerning partitions. 

In section \ref{sec:isomorphism} we define the isomorphism $\Omega$ which was discussed in section 1.1. Actually we define an infinite family of isomorphisms  $ \{ \Omega_K \: | \: K \in \zint \} $ between the space of states $\F$ and the space of symmetric Laurent polynomials in variables $x_1,\dots,x_n .$ The isomorphisms $\Omega_K$ are related to each other by the trivial shift:    
\begin{equation}
 \Omega_K =  (x_1 \dots x_n) \Omega_{K+1}
\end{equation}
and the isomorphism $\Omega$ whose existence is claimed in section 1.1 is $ \Omega_K$ with an arbitrary integer $K.$ The main part of  this section is the proof that each of the $ \Omega_K$ is an isomorphism of {\em Hilbert spaces} i.e. that it respects scalar products.

In the brief section \ref{sec:basis} we describe  the basis in the space of symmetric Laurent polynomials obtained from the  $A(\glN;\beta)$-eigenbasis by the map with the isomorphism $ \Omega_K$ for any fixed $K.$ 

In section \ref{sec:jack} we define the $\glN$-Jack Polynomials and discuss some of their properties.

In section \ref{sec:identification} we establish the main result of this paper which was described in section 1.1.

Finally, in sections \ref{sec:parameterization} and \ref{sec:correlation} we compute the Correlation Functions.
The Appendix contains proofs of some of the statements in the main text. 
\\ \mbox{} 

\noindent {\bf {\Large Acknowledgments }} I am grateful to Kouichi Takemura with whom we collaborated  on the paper  \cite{TU} from which the present work has evolved and  to Professor P.J. Forrester for encouragement and discussions during his stay at RIMS, Kyoto University in Summer 1996. I am also grateful to Professors T. Miwa and M. Kashiwara for discussions and support. 

After this work had been completed I have learned about the unpublished typescript of Y. Kato \cite{Kato} in which he computes the Green Function of the Spin Calogero-Sutherland Model for $N=2$ and $\beta = 1$ and conjectures expressions for the Green Function at all integer values of $\beta.$ The method of Kato is completely different from the method of our paper and is in the spirit of the papers \cite{KatoKu, Forrester}. I am grateful to T. Yamamoto for bringing the paper  \cite{Kato} to my attention.        

\section{Gauge-transformed Hamiltonian of the Spin Calogero-Sutherland Model}
In this part of the paper we summarize some properties of the gauge-transformed Hamiltonian $H_{\beta,N}$ and give the definition of the Hilbert space of states on which it acts. This part mainly follows the paper \cite{BGHP} and our primary objective here is to introduce our notations and to formulate definition of the Model in a way  suitable for  our subsequent considerations.  

\subsection{Preliminary remarks and notations} \label{sec:preliminary}
Let $N$ be a positive integer. In this paper $N$ has the meaning of the  number of spin degrees of freedom of each particle in the Spin Calogero-Sutherland Model. For any integer $k$ define the unique $\un{k}$ $\in$ $\setN$ and the unique  $\ov{k}$ $\in$ $\zint$ by setting $k = \un{k} - N \ov{k}$. And for a $k$ $=$ $(k_1,k_2,\dots,k_n)$ $\in$ $\zint^n$ set $\un{k}$ $=$ $(\un{k_1},\un{k_2},\dots,\un{k_n})$, $\ov{k}$ $=$ $(\ov{k_1},\ov{k_2},\dots,\ov{k_n}).$ 

 For any sequence $k$ $=$ $(k_1,k_2,\dots,k_n)$ $\in$ $\zint^n$ let $|k|$ be the  weight: $|k|$ $=$ $k_1 + k_2 + \cdots + k_n$, and define the partial ordering ( the {\em natural} or the {\em dominance } ordering \cite{MacBook} ) on $\zint^n$ by setting for any two distinct $k,l$ $\in$ $\zint^n$:   
\begin{gather} k > l  \\ 
\text{ iff } \quad |k| = |l|, \quad \text{ and} \quad  k_1+\cdots+k_i \geq l_1+\cdots+l_i\quad \text{ for all} \quad i=1,2,\dots,n. \nonumber 
\end{gather}
For $r\in \nat$ let $\LC^{(r)}_n$ be a subset of $\zint^n$ defined as
\begin{equation}
\LC^{(r)}_n = \{ k = (k_1,k_2,\dots,k_n) \in \zint^n \; | \; k_i \geq k_{i+1} \quad \text{and} \quad \forall s \in \zint \; \#\{k_i \: | \: k_i = s\} \leq r \}. \label{eq:LC}
\end{equation}
In particular the $\LC^{(n)}_n$ is the set of non-increasing sequences of $n$ integers and   the $\LC^{(1)}_n$ is the set of {\em strictly decreasing } sequences, i.e. such $k$ $ = $ $ (k_1,k_2,\dots,k_n) \in \zint^n$ that $ k_i > k_{i+1}$.    

Let $V$ = $\cplxN$ with the basis $\{ v_1,v_2,\dots,v_N\}$ and let $V(z)$ $=$ $ \cplx[z^{\pm 1}]\otimes V $ with the basis $\{ u_k\:|\: k\in \zint\}$ where $   u_k = z^{\ov{k}}\otimes v_{\un{k}}.$ For monomials in vector spaces $\cz$, $\otimes^n V$ and $\otimes^n V(z)$ $=$ $\cz\otimes(\otimes^n V)$ we will use the convention of multi-indices:  
\begin{eqnarray}
&z^t  =  z_1^{t_1}z_2^{t_2}\cdots z_n^{t_n}, \qquad & t=(t_1,t_2,\dots,t_n) \in \zint^n; \\
&v(a) =  v_{a_1}\otimes v_{a_2}\otimes\cdots \otimes v_{a_n},\qquad &  a=(a_1,a_2,\dots,a_n) \in \setN^n;  \\ 
&u_k  =   u_{k_1}\otimes u_{k_2}\otimes \cdots \otimes u_{k_n}, \qquad & k = (k_1,k_2,\dots,k_n) \in \zint^n. 
\end{eqnarray}
Let $K_{ij}$ be the permutation operator for variables $z_i$ and $z_j$ in $\cz$ ( operator of coordinate permutation ), and let $P_{ij}$ be the operator exchanging $i$th and $j$th factors in the tensor product $\otimes^n V$ ( operator of spin permutation ).

Let $A_n$ be the antisymmetrization operator in $\otimes^n V(z)$: 
\begin{equation}
A_n (u_{k_1}\otimes u_{k_2}\otimes \cdots \otimes u_{k_n}) = \sum_{w\in S_n}\sgn(w) u_{k_{w(1)}}\otimes u_{k_{w(2)}}\otimes \cdots \otimes u_{k_{w(n)}}, \label{eq:antisymm}
\end{equation}
where $S_n$ is the symmetric group of order $n$. We will  use the notation $\hat{u}_k$ $=$ $u_{k_1}\wedge u_{k_2}\wedge \cdots \wedge u_{k_n} $ for a vector of the form (\ref{eq:antisymm}), and will  call such a vector {\em a wedge}. A wedge $\hat{u}_k$ is {\em normally ordered } if $k \in \LC^{(1)}_n$, that is $k_1>k_2>\dots >k_n$. Let $F_{N,n}$ be the image of the operator $A_n$ in $\otimes^n V(z)$. Then $F_{N,n}$ is spanned by  wedges and the normally ordered wedges form a basis in $F_{N,n}$. Equivalently the vector space $\F$ is defined as the linear span of all vectors $f$ $\in$ $\cz \otimes (\otimes^n V) $ such that for all $1\leq i\neq j \leq n$:  
\begin{equation}
                   K_{ij}f = -P_{ij}f. 
\end{equation}
This is the definition of $\F$ adopted in \cite{BGHP}.

\subsection{Scalar product} \label{sec:scalar}
Here we define a scalar product on the space of states $\F.$ Our definition has three  steps. Frst we define scalar products on the vector spaces $\otimes^n V$ and $\cz$ separately. Then we define a scalar product on the tensor product $\otimes^n V(z)$ $=$ $\cz \otimes ( \otimes^n V).$ Finally we define a scalar product on $\F$ considered as a subspace of $\otimes^n V(z).$  

On $\otimes^n V$ define a sesquilinear, i.e., anti-linear in the first argument and linear in the second argument,  scalar product ${\sprod{\cdot}{\cdot}}_N$ by requiring that pure tensors in $\otimes^n V$ be orthonormal: 
\begin{equation}
 {\sprod{v(a)}{v(b)}}_N = \delta_{ab}, \qquad a,b \in \setN^n.
\end{equation}
For $ w=(w_1,w_2,\dots,w_n) \in \cplx^n$, $|w_1|=|w_2|=\dots=|w_n|=1$ and a non-negative real number $\delta$ let:
\begin{equation}
\Delta(w;\delta) = \prod_{1\leq i\neq j \leq n}(1 - w_i w_j^{-1})^{\delta},
\end{equation}
and define for all $f(z),g(z)$ $\in$  $\cz$ a scalar product ${\sprod{\cdot}{\cdot}}_{\delta}'$ by setting :
\begin{equation}
{\sprod{f(z)}{g(z)}}_{\delta}' = \frac{1}{n!}\prod_{j=1}^n \int \frac{dw_j}{2\pi i w_j} \Delta(w;\delta) \ov{f(w)}g(w) , \label{eq:spc}
\end{equation}
where the integration over each of the variables $w_j$ is taken along the unit circle in the complex plane, and the bar over $f(w)$ means complex conjugation.

On the linear space $\otimes^n V(z)$ $=$ $\cz \otimes (\otimes^n V)$ we define a scalar product ${\sprod{\cdot}{\cdot}}_{\delta,N}'$ as the composition of the scalar products ${\sprod{\cdot}{\cdot}}_{\delta}'$ and ${\sprod{\cdot}{\cdot}}_N$, i.e. for  $f(z),g(z)$ $\in$ $\cz ;$ $u,v$ $\in$ $\otimes^n V$ we set:  
\begin{equation}
{\sprod{f(z)\otimes u}{g(z)\otimes v}}_{\delta,N}' ={\sprod{f(z)}{g(z)}}_{\delta}' {\sprod{u}{ v}}_{N},\label{eq:spcv}
\end{equation}
and extend the definition on all vectors by requiring that ${\sprod{\cdot}{\cdot}}_{\delta,N}'$ be sesquilinear.

Finally, on the subspace $\F \subset \otimes^n V(z)$ a scalar product  ${\sprod{\cdot}{\cdot}}_{\delta,N}$ is defined as the restriction of  the scalar product ${\sprod{\cdot}{\cdot}}_{\delta,N}'$. Note, that the normally ordered wedges are orthonormal relative to this scalar product when $\delta =0$: 
\begin{equation}
{\sprod{ \hat{u}_k }{\hat{u}_l}}_{0,N} = \delta_{kl}, \qquad k,l \in \LC^{(1)}_n.
\end{equation}

\subsection{The gauge-transformed Hamiltonian} \label{sec:gauge}
Here we briefly recall the relationship between the gauge-transformed Hamiltonian and the Dunkl operators \cite{BGHP}.

We will use the following convention: if $B$ is an operator acting on $\cz$ ( or $\otimes^n V$ ) then we will denote by the same letter $B$ the operator $B\otimes\id$ ( or $\id\otimes B$ ) acting on the space $\cz\otimes(\otimes^n V ).$

Let $W = \prod_{1\leq i<j\leq n}\sin\frac{\pi}{L}(y_i - y_j)$ and set $z_j = \exp(\frac{2\pi i}{L}y_j).$ Then the gauge-transformed Hamiltonian $H_{\beta,N}$ is defined as follows \cite{BGHP}:
\begin{gather}
H_{\beta,N} = \frac{L^2}{2\pi^2} W^{-\beta} \ov{H}_{\beta,N} W^{\beta} = \label{eq:H}\\ 
= \sum_{i=1}^n D_i^2 + \beta\sum_{i=1}^n(2i-n-1)D_i + 2\beta\sum_{1\leq i<j\leq n}\theta_{ij}\left(D_i - D_j + \theta_{ji}(P_{ij} + 1)\right) + \frac{\beta^2n(n^2-1)}{12}, \nonumber 
\end{gather}
where we set: $D_i = z_i{\p}/{\p z_i}$ and $\theta_{ij}=z_i/(z_i-z_j)$. 

The Hamiltonian $H_{\beta,N}$ acts in the space $\F$ \cite{BGHP}. To see this let us recall, that the Dunkl operators \cite{Dunkl}: 
\begin{equation}
d_i(\beta)=\beta^{-1}D_i - i + \sum_{i<j}\theta_{ji}(K_{ij}-1)-\sum_{i>j}\theta_{ij}(K_{ij}-1), \qquad (i=1,2,\dots,n),
\end{equation}
act on $\cz$ and together with permutations $K_{ij}$ satisfy the defining relations of the degenerate affine Hecke algebra:
\begin{align}
& K_{ii+1}d_i(\beta) - d_{i+1}(\beta)K_{ii+1} = 1, \label{eq:hecke1}\\ 
& K_{ii+1}d_j(\beta) = d_j(\beta)K_{ii+1}, \quad (j\neq i,i+1), \label{eq:hecke2}\\ 
& d_i(\beta)d_j(\beta) = d_j(\beta)d_i(\beta). \label{eq:hecke3}
\end{align}
These relations imply, in particular, that symmetric polynomials in Dunkl operators commute with permutations $K_{ij}$ and therefore are well-defined operators on the space $\F$. The action of the operator
\begin{gather}
\beta^2\sum_{i=1}^n \left( d_i(\beta) + \frac{n+1}{2}\right)^2 = \label{eq:d=H}\\
=\sum_{i=1}^n D_i^2 + \beta\sum_{i=1}^n(2i-n-1)D_i + 2\beta\sum_{1\leq i<j\leq n}\theta_{ij}\left(D_i - D_j - \theta_{ji}(K_{ij} - 1)\right) + \frac{\beta^2n(n^2-1)}{12} \nonumber 
\end{gather}
on any vector $f\in \F$ is seen to coincide with the action of the Hamiltonian $H_{\beta,N}$ since $K_{ij}f=-P_{ij}f$. Thus we may write:
\begin{equation}
H_{\beta,N} = \beta^2\sum_{i=1}^n \left( d_i(\beta) + \frac{n+1}{2}\right)^2
\end{equation}  \label{eq:H=d}
as operators on $\F$.

Note that from the last equation it follows  that the $H_{\beta,N}$ is self-adjoint relative to the scalar product ${\sprod{\cdot}{\cdot}}_{\beta,N}$ since the Dunkl operators are self-adjoint relative to the scalar product ${\sprod{\cdot}{\cdot}}_{\beta}'$ as one can easily verify.

\subsection{An eigenbasis of the Hamiltonian $H_{\beta,N}$} \label{sec:eigenbasis1}

In this section we construct a certain eigenbasis of the Hamiltonian $H_{\beta,N}.$ This eigenbasis is not orthogonal with respect to the scalar product ${\sprod{\cdot}{\cdot}}_{\beta,N}.$ However it has a {\em doubly} triangular expansion in the basis of wedges. This is an important property to be used later in section \ref{sec:a-eigenbasis}.   

Let $k=(k_1,k_2,\dots,k_n)\in \LC^{(1)}_n$. Note that $k_1 > k_2 > \dots > k_n$ implies, in particular, that $\ov{k_1}\leq \ov{k_2}\leq \dots \leq \ov{k_n}$. And let us act with the Hamiltonian $H_{\beta,N}$ on the (normally  ordered) wedge $\hat{u}_k$ $=$ $u_{k_1}\wedge u_{k_2} \wedge \dots \wedge u_{k_n}$. Using the expression (\ref{eq:d=H}) we find: 
\begin{equation}
H_{\beta,N}\hat{u}_k = E(k;\beta)\hat{u}_k + 2\beta\sum_{1\leq i<j\leq n}h_{ij}\hat{u}_k, \label{eq:Hw=}
\end{equation}
where 
\begin{gather}
E(k;\beta)=\sum_{i=1}^n \ov{k_i}^2 + \beta\sum_{i=1}^n(2i-n-1)\ov{k_i} + \frac{\beta^2n(n^2-1)}{12}, \label{eq:Heigenvalue} \\
\text{and} \quad h_{ij}(u_{k_1}\wedge\dots\wedge u_{k_i}\wedge \dots\wedge u_{k_j} \wedge \dots \wedge u_{k_n}) =\label{eq:hij} \\ = \sum_{r=1}^{\ov{k_j}-\ov{k_i}-1}(\ov{k_j}-\ov{k_i}-r)(u_{k_1}\wedge\dots\wedge u_{k_i-Nr}\wedge \dots\wedge u_{k_j+Nr} \wedge \dots \wedge u_{k_n}). \nonumber 
\end{gather}
Normally ordering the wedges in the right-hand side of (\ref{eq:Hw=}) we find from (\ref{eq:hij}) that: 
\begin{equation}
H_{\beta,N}\hat{u}_k = E(k;\beta)\hat{u}_k + \sum\begin{Sb} l \in \LC^{(1)}_n,\; 
\ov{l} > \ov{k}, \; l < k \end{Sb} h_{kl}^{(\beta)} \hat{u}_l, \label{eq:triang1}
\end{equation}
with certain real coefficients  $h_{kl}^{(\beta)}$.

Now recall from \cite{MacBook} that for positive $\beta$ we have: $ E(k;\beta)\neq E(l;\beta)$ when $\ov{l} > \ov{k}$. Then (\ref{eq:triang1}) leads to:
\begin{em}
\begin{align}
& \text{For any $k\in\LC^{(1)}_n$ there is a unique eigenvector $\Psi_k^{(\beta)}$ of $H_{\beta,N}$ such that:} \label{p1i} \\ 
& \Psi_k^{(\beta)} =\hat{u}_k + \sum\begin{Sb} l \in \LC^{(1)}_n,\; 
\ov{l} > \ov{k} \end{Sb} \psi_{kl}^{(\beta)} \hat{u}_l  \qquad ( \psi_{kl}^{(\beta)}  \in \real ). \nonumber \\
& \text{Eigenvalue of $H_{\beta,N}$ for this eigenvector is $ E(k;\beta).$} \label{p1ii} \\
& \text{The coefficient  $\psi_{kl}^{(\beta)}$ vanishes unless $l < k.$}
\label{p1iii}
\end{align}
\end{em}
Note that for any integer $M$ the wedge:
\begin{equation}
\vac(M) = u_M\wedge u_{M-1}\wedge \cdots \wedge u_{M-n+1} \label{eq:vac}
\end{equation}
is an eigenvector of the Hamiltonian $H_{\beta,N}$ as implied by either (\ref{p1i}) or (\ref{eq:Hw=}, \ref{eq:hij}). We will call any vector of the form (\ref{eq:vac}) a {\em vacuum } vector. 

As we have mentioned already, the eigenbasis $\{ \Psi_k^{(\beta)} \}$ is not orthogonal. To construct an orthogonal eigenbasis we need to utilize the Yangian symmetry of the Model as discussed in the next part of the paper.

\section{Yangian Gelfand-Zetlin bases and an orthogonal eigenbasis of the Spin Calogero-Sutherland Model}
Our objective in this part of the paper is to construct an orthogonal eigenbasis of the Spin Calogero-Sutherland Model. As was shown in the work \cite{TU} to do this it is natural to use the Yangian symmetry of the Model. The orthogonal eigenbasis then is defined uniquely up to normalization as the eigenbasis of the commutative family of operators which give the action of the maximal commutative subalgebra in the Yangian action on the space of states $\F.$

\subsection{The Yangian of $\glN$ and its maximal commutative subalgebra} \label{sec:yangian}
The Yangian $Y(\glN)$ \cite{Drinfeld} is an associative unital algebra with generators: the unit $1$ and $T^{(s)}_{ab}$ where $a,b \in \setN$ and  $s=1,2,\dots\; .$ In terms of the formal power series in variable $u^{-1}$:
\begin{equation}
T_{ab}(u) = \delta_{ab}1 + u^{-1}T_{ab}^{(1)} +  u^{-2}T_{ab}^{(2)} + \cdots 
\end{equation}
the relations of $Y(\glN)$ are written as follows
\begin{equation}
(u - v)[T_{ab}(u), T_{cd}(v)] = T_{cb}(v)T_{ad}(u) - T_{cb}(u)T_{ad}(v),
\end{equation}
and the coproduct $\Delta$ $:$ $Y(\glN)$ $\rightarrow$ $Y(\glN)\otimes Y(\glN)$
is given by: $\Delta(T_{ab}(u))$ $=$ $\sum_{c=1}^{N} T_{ac}(u)\otimes T_{cb}(u).$

The center of $Y(\glN)$  is generated by coefficients of the series
\begin{equation}
A_N(u) = \sum_{w\in S_N}\sgn(w) T_{1 w(1)}(u)T_{2 w(2)}(u-1) \cdots T_{N w(N)}(u-N+1)
\end{equation}
called the Quantum Determinant of $Y(\glN)$. 

The Yangian has a distinguished maximal commutative subalgebra $A(\glN)$ \cite{Cherednik,NT}. This subalgebra is generated by coefficients of the series $A_1(u),A_2(u),\dots,A_N(u)$ defined as follows:  
\begin{equation}
A_m(u) = \sum_{w\in S_m}\sgn(w) T_{1 w(1)}(u)T_{2 w(2)}(u-1) \cdots T_{m w(m)}(u-m+1), \qquad m=1,2,\dots,N.
\end{equation}
That is to say by the centers of all algebras in the chain:
\begin{equation}
 Y(\gl_1) \subset Y(\gl_2) \subset \dots \subset Y(\gl_N), 
\end{equation}
where for $m=1,2,\dots,N-1$ the $Y(\gl_m)$ is realized inside  $Y(\gl_N)$ as the subalgebra generated by coefficients of the series $T_{ab}(u)$ with $a,b =1,2,\dots,m. $

Finite or infinite dimensional $Y(\glN)$-modules with a semisimple action of the subalgebra $A(\glN)$ are called {\em tame}, and eigenbases of $A(\glN)$ in irreducible tame modules are called ( Yangian ) {\em Gelfand-Zetlin bases } \cite{NT}.

Let us now consider certain tame Yangian modules which appear in the context of the Fermionic Spin Calogero-Sutherland Model. Let $f\in \cplx$ and $\pi(f)$ $:$ $Y(\gl_N)$ $\rightarrow$ $End(V)$ be the $Y(\gl_N)$-homomorphism defined by:
\begin{equation}    
\pi(f)(T_{ab}(u)) = \delta_{ab}1 + \frac{E_{ba}}{u + f}, \label{eq:evhom}
\end{equation}
where $E_{ba}$ $\in$ $End(V)$ is the matrix unit: $ E_{ba}v_c = \delta_{ac}v_b.$ Also for $f=(f_1,f_2,\dots,f_n) \in \cplx^n$ we will denote by $\pi(f)$ the tensor product of the homomorphisms (\ref{eq:evhom}): $\pi(f) = $ $ \pi(f_1)\otimes\pi(f_2)\otimes \cdots \otimes \pi(f_n)$ $:$ $\otimes^n Y(\glN)$ $\rightarrow$ $End(\otimes^n V),$ so that $ \pi(f)\left(\Delta^{(n)}(T_{ab}(u))\right), $ where $ \Delta^{(n)}$ is the coproduct iterated $n-1$-times, defines a $Y(\glN)$-module structure on $\otimes^n V$.  

Let now $M$ be an integer ( unrelated to the $M$  in (\ref{eq:vac})) and let $p = (p_1,p_2,\dots,p_M) \in \setN^M$ be a sequence of integers such that:
\begin{equation}
n = p_1+p_2+\cdots + p_M. \label{eq:pseq}
\end{equation}
With these $(p_1,p_2,\dots,p_M)$ define integers $ q_0,q_1,\dots,q_M $ by 
\begin{equation}
p_s = q_s - q_{s-1}, \quad (q_0 := 0), \qquad ( s=1,2,\dots,M ). \label{eq:p=q-q}
\end{equation}
For $ 1 \leq i < j \leq n $ define the partial anti-symmetrization operator  
$A_{(i,j)}$ $\in$ $End(\otimes^n V)$ by setting for $a$ $=$ $(a_1,\dots,a_n)$ $ \in \setN^n$: 
\begin{multline}
A_{(i,j)}(v_{a_1}\otimes v_{a_2}\otimes\cdots\otimes v_{a_n}) := \\ \sum_{w\in S_{j-i}}\sgn(w) v_{a_1}\otimes v_{a_2}\otimes\cdots\otimes v_{a_i}\otimes v_{a_{i+w(1)}}\otimes v_{a_{i+w(2)}}\otimes \cdots \otimes v_{a_{i+w(j-i)}}\otimes v_{a_{j+1}}\otimes \cdots   \otimes v_{a_n}. 
\end{multline} 
And let $(\otimes^n V )_{p}$ be the image in $\otimes^n V $ of the operator:
\begin{equation}
 A_p := \prod_{s=1}^M A_{(q_{s-1},q_s)}.  
\end{equation}
If we define the  set $T_p$ labeled by the sequence $p$ by   
\begin{multline}
T_p :=  \{ a = (a_1,\dots,a_n ) \in \setN^n \; | \; a_i < a_{i+1} \; \text{when}\; q_{s-1} < i < q_s \; \text{ for each } \; s=1,2,\dots, M \} \end{multline}
then the set 
\begin{equation}
\{ \varphi(a):= A_p v(a) \; | \; a \in T_p \}  \label{eq:phi-def}
\end{equation}
is a basis of $ (\otimes^n V )_{p}$.

Further, let $ f =(f_1,{f}_2,\dots,{f}_n) $ be a sequence of real numbers satisfying  the following two conditions:
\begin{align}
& \text{ ${f}_i = {f}_{i+1}+1 $ when $ q_{s-1} < i < q_s $ for each $  s=1,2,\dots, M; $  }  \tag{C1}\label{eq:fbar1} \\
& \text{and} \qquad {f}_{q_s} > {f}_{q_s+1} + 1   \qquad \text{for} \quad   s=1,2,\dots, M-1.  \tag{C2} \label{eq:fbar2}
\end{align}
Then we have:
\begin{em}
\begin{align}
& \text{The coefficients of $ \pi(f)\Delta^{(n)}( T_{ab}(u) ) $ $\in$ $End(\otimes^n V)[[u^{-1}]]$ }\label{p2i}\\ 
& \text{leave  the subspace  $(\otimes^n V )_{p} \subset \otimes^n V$  invariant.}  \nonumber \\
& \text{In $( \otimes^n V)_{p}$ there is a unique up to normalization of eigenvectors}\label{p2ii} \\ &  \pi(f)A(\glN)-\text{eigenbasis:} \quad \{ \chi(a) \; | \; a \in T_p \}. \nonumber   \\  
& \chi(a) = \varphi(a) + \sum_{b > a} c(a,b) \varphi(b), \qquad c(a,b) \in \real. \label{p2iii} \\ 
& \pi(f) \Delta^{(n)}( A_m(u) ) \chi(a) = A_m(u;f;a) \chi(a), \qquad m=1,2,\dots,N; \label{p2iv}\\  
& \text{where} \quad  A_m(u;f;a):= \prod_{i=1}^n \frac{u+f_i + \delta(a_i \leq m) }{u+ f_i}. \nonumber \\ 
& \text{The $N$-tuples of rational functions in $u$: $  A_1(u;f;a), \dots,  A_N(u;f;a) $ } \label{p2v} \\  
& \text{are distinct for distinct $a \in T_p$.  In other words: } \nonumber \\ 
& \text{the spectrum of the  $\pi(f)A(\glN)$ on $(\otimes^n V)_p$ is simple.}\nonumber   
\end{align}
\end{em}
In the expression for the eigenvalue $A_m(u;f;a)$ above we have used the convention that for a statement $P,$ $\delta(P) = 1$ if $P$ is true, and $\delta(P) = 0$ otherwise. We will often use this convention in this paper.
We give  proofs of the statements (\ref{p2i}) - (\ref{p2v}) in Appendix, section A.

\subsection{Yangian in the Spin Calogero-Sutherland Model} \label{sec:yangian in scsm}
The space $\F$ admits a $Y(\glN)$-action defined as follows \cite{BGHP}: let $E_{ab}^{(i)}$ $=$ $1^{\otimes^{i-1}}\otimes E_{ab}\otimes 1^{\otimes^{n-i}}$ $\in$ $End(\otimes^n V),$ and let 
\begin{equation}
L_{ab}^{(i)}(u;\beta) = \delta_{ab}1 + \frac{E_{ab}^{(i)}}{u + d_i(\beta)}. 
\end{equation}
Set $T_{ab}(u;\beta) = L_{ac_1}^{(1)}(u;\beta)L_{c_1c_2}^{(2)}(u;\beta)\cdots L_{c_{n-1}b}^{(n)}(u;\beta),$ where summation over the indices $ c_1,\dots, c_{n-1} $ is assumed. The degenerate affine Hecke algebra relations satisfied by the Dunkl operators imply that coefficients of the series $T_{ab}(u;\beta)$ act on $\F$ and satsify the defining relations of the Yangian \cite{BGHP}. Denote by $Y(\glN;\beta)$ the $Y(\glN)$-action on $\F$ defined by the   $T_{ab}(u;\beta)$ and denote by $A(\glN;\beta)$ the corresponding action of the subalgebra $A(\glN)$. In particular the Quantum Determinant of $T_{ab}(u;\beta)$ has the form \cite{BGHP}:  
\begin{equation}
A_N(u;\beta) = \prod_{i=1}^n\frac{ u + 1 + d_i(\beta)}{u + d_i(\beta)}, 
\end{equation}
and therefore the Hamiltonian $H_{\beta,N}$ (\ref{eq:H=d}) is an element in the center of the action $Y(\glN;\beta)$ and hence an element in $A(\glN;\beta)$.

For an operator $B$ acting on $\F$ let $B^*$ be its adjoint relative to the scalar product ${\sprod{\cdot}{\cdot} }_{\beta,N}$ and for a series $B(u)$ with operator-valued coefficients we will use $B(u)^*$ to denote the series whose coefficients are adjoints of coefficients of the series $B(u)$. 

In \cite{TU} it is shown, that generators of the action  $Y(\glN;\beta)$ satisfy the following relations:
\begin{equation}
T_{ab}(u;\beta)^* = T_{ba}(u;\beta).
\end{equation}
These relations imply, in particular, that the action of the subalgebra $A(\glN;\beta)$ is self-adjoint. That is we have \cite{TU}:  
\begin{equation}
A_m(u;\beta)^* = A_m(u;\beta), \qquad m=1,2,\dots,N. \label{eq:selfa}
\end{equation}
 
\subsection{Decomposition of the space of states with respect to the Yangian action 
$Y(\glN;\beta)$} \label{sec:decomposition}
Here we recall the decomposition of the space of states $\F$ into irreducible Yangian representations with respect to the Yangian action $Y(\glN;\beta).$ For details one may consult the work \cite{TU} which contains  proofs of the statements made in this section. 

The Yangian decomposition of the space of states $\F$ is constructed by using the Non-symmetric Jack Polynomials ( see \cite{Opdam, Cherednik2, Macdonald1} ).

Non-symmetric Jack polynomials $ E_t^{(\beta)}(z), $ $t = (t_1,\dots,t_n) \in \zint^n$ form a basis of $\cz$. For any $t \in \zint^n$ let $t^+$ denote the element of the set $\LC_n^{(n)}$ (\ref{eq:LC}) obtained by arranging parts of $t$ in non-increasing order. The polynomials $ E_t^{(\beta)}(z)$ have the triangular expansion in the monomial basis:
\begin{equation}
 E_t^{(\beta)}(z) = z^t + \sum_{r \prec t}e^{(\beta)}_{tr} z^r, \label{eq:nsJtri}
\end{equation}
where $e^{(\beta)}_{tr}$ are real coefficients and
\begin{equation}
r \prec t \quad \Leftrightarrow \quad \begin{cases} t^+ > r^+  &  \text{or} \\ 
   t^+ = r^+  & \text{and the last non-zero difference $t_i - r_i$ is negative.}\end{cases} \nonumber 
\end{equation}
Moreover, polynomials $ E_t^{(\beta)}(z)$  are eigenvectors of the Dunkl operators: 
\begin{align}
& d_i(\beta)E_t^{(\beta)}(z) = f_i(t;\beta)E_t^{(\beta)}(z), \qquad (i=1,2,\dots,n),  \\ 
& \text{where} \quad f_i(t;\beta) = \beta^{-1}t_i - \rho_i(t), \label{eq:d-eigenvalue}\\ 
& \text{and} \quad \rho_i(t) = \#\{ j \leq i \: | \: t_j \geq t_i \} + \#\{ j > i \: | \: t_j > t_i \}.
\end{align}
In \cite{TU} it is shown, that the space $\F$ splits into an infinite number of irreducible Yangian submodules $ F_s$ labelled by elements of the set $\LC_n^{(N)}$ (\ref{eq:LC}):
\begin{equation}
\F = \oplus_{s \in \LC^{(N)}_n} F_s. \label{eq:Yangdecomp}
\end{equation}
To describe  the component $F_s$ which corresponds to a given $ s\in\LC_n^{(N)}$ let $M$ be the number of distinct elements in the sequence $s$ $=$ $(s_1,s_2,\dots,s_n)$. And let $q_0,q_1,\dots,q_M$ be defined by: $q_0 = 0, q_M = n;$ $s_{q_j} >  s_{q_j+1}$, $ j=1,2,\dots,M-1.$ As in (\ref{eq:p=q-q}) a sequence $p(s)=(p_1,p_2,\dots,p_M)$ is defined by: $ p_j = q_j - q_{j-1},$ $ j=1,2,\dots,M$. Clearly we have $ p_1+p_2+ \cdots +p_M$ $=$ $n.$     

As a $Y(\glN)$-module the space $F_s$ is isomorphic to $(\otimes^n V)_{p(s)}$ (\ref{sec:yangian}) where the Yangian action is given by $ \pi(f(s))\Delta^{(n)}(T_{ab}(u))$ and $ f(s) = (f_1,f_2,\dots,f_n)$ with $ f_i = f_i(s;\beta).$ This isomorphism is explicitly given by the operator $U(s;\beta):$ $(\otimes^n V)_{p(s)}$ $\rightarrow$ $ F_s$ which is defined for any $v\in (\otimes^n V)_{p(s)}$ as follows:   
\begin{equation}
U(s;\beta)v = \sum_{t \sim s} E_t^{(\beta)}(z)\otimes R_t^{(\beta)}v,   \label{eq:Uop}
\end{equation}
where the sum is taken over all distinct rearrangements $t$ of $s$, and $ R_t^{(\beta)}$ $ \in$ $End(\otimes^n V)$ is defined by the recursive realtions:  
\begin{gather}
 R_s^{(\beta)} = 1, \\ 
 R_{t(i,i+1)}^{(\beta)}= -\check{R}_{ii+1}( f_i(t;\beta) - f_{i+1}(t;\beta))R_{t}^{(\beta)} \quad \text{for} \quad t_i > t_{i+1}. 
\end{gather}
Here $\check{R}_{ii+1}(u)$ $=$ $u^{-1} + P_{ii+1}$ and $t(i,i+1)$ denotes the element of $\zint^n$ obtained by interchanging $t_i$ and $t_{i+1}$ in $t$ $=$ $(t_1,\dots,t_n)$.  

Observe now that, since  $\beta >0$, the conditions  (\ref{eq:fbar1},\ref{eq:fbar2}) are satisfied by the sequence $ f(s)$. Therefore setting for all $a \in T_{p(s)}$:   
\begin{eqnarray} 
 \Phi^{(\beta)}(s,a) & := & U(s;\beta)\varphi(a), \\ 
 X^{(\beta)}(s,a) & := & U(s;\beta)\chi(a);
\end{eqnarray}
from (\ref{p2ii} - \ref{p2v}) and the fact that $F_s$ is isomorphic to $(\otimes^n V)_{p(s)}$ we obtain the following: 

\begin{em}
\begin{align}
& \text{$\{ X^{(\beta)}(s,a) \; | \; a \in T_{p(s)} \}$ is the  unique up to normalization of eigenvectors}\label{p3i} \\ &  A(\glN;\beta)-\text{eigenbasis of $F_s.$} \nonumber \\
&  X^{(\beta)}(s,a) = \Phi^{(\beta)}(s,a) + \sum_{b > a} c(a,b) \Phi^{(\beta)}(s,b), \qquad c(a,b) \in \real. \label{p3ii} \\ 
&  A_m(u;\beta) X^{(\beta)}(s,a) = A_m(u;f(s);a)X^{(\beta)}(s,a), \qquad m=1,2,\dots,N; \label{p3iii}\\  
& \text{where} \quad  A_m(u;f(s);a):= \prod_{i=1}^n \frac{u+f_i(s;\beta) + \delta(a_i \leq m) }{u+ f_i(s;\beta)}. \nonumber \\ 
& \text{The $N$-tuples of rational functions in $u$: $  A_1(u;f(s);a), \dots,  A_N(u;f(s);a) $ } \label{p3iv} \\  
& \text{are distinct for distinct $a \in T_{p(s)}$.} \nonumber 
\end{align}
\end{em}

Thus the set $\{ X^{(\beta)}(s,a) \; | \; a \in T_{p(s)} \}$ is a Yangian Gelfand-Zetlin basis of the irreducible Yangian representation $F_{s}. $

\subsection{$A(\glN;\beta)$-eigenbasis of the space of states $\F$} \label{sec:a-eigenbasis}
Here we define the orthogonal eigenbasis of the commutative family $A(\glN;\beta)$ in the entire space of states $\F.$ This eigenbasis is just the union of bases $\{ X^{(\beta)}(s,a) \; | \; a \in T_{p(s)} \}$ taken over all elements $s$ from the set $\LC_n^{(N)}.$ We prefer, however, to choose a different parameterization of elements in this basis, such that a triangularity of these elements expanded in the basis of normally ordered wedges becomes manifest.

For any $k = (k_1,k_2,\dots,k_n) \in \LC_n^{(1)}$ the sequence $s$ $  =$ $  (\ov{k_n},\ov{k_{n-1}},\dots,\ov{k_1})$ is an element of the set $\LC_n^{(N)}$. And the sequence $ a $ $ = $ $(\un{k_n},\un{k_{n-1}},\dots,\un{k_1})$ belongs to the set $T_{p(s)}$ defined by the sequence $s$ as in the previous section. With these $k,$ $s$ and $a$ we have:   
\begin{equation}
(-1)^{\frac{n(n-1)}{2}}\Phi^{(\beta)}(s,a) = \Psi_k^{(\beta)}, \label{eq:Phi=Psi}
\end{equation}
where the $\Psi_k^{(\beta)}$ is the eigenvector of the Hamiltonian $H_{\beta,N}$ defined in (\ref{p1i}). A proof of the equality  (\ref{eq:Phi=Psi}) is contained in the Appendix, section B.

Now with the same  $k,$ $s$ and $a$ as above let us define:
\begin{equation} 
X_k^{(\beta,N)} := (-1)^{\frac{n(n-1)}{2}}X^{(\beta)}(s,a).
\end{equation}
Then we have the following statements about the vectors $ X_k^{(\beta,N)}$:
\begin{em}
\begin{align}
& \text{$\{ X_k^{(\beta,N)} \; | \; k \in \LC_n^{(1)} \}$ is the  unique up to normalization of eigenvectors}\label{p4i} \\ &  A(\glN;\beta)-\text{eigenbasis of $\F.$} \nonumber \\
&  X_k^{(\beta,N)} = \hat{u}_k  + \sum_{l \in \LC_n^{(1)}, \; l < k,\; \ov{l} \geq \ov{k}} x^{(\beta)}_{kl} \;\hat{u}_l, \qquad  x^{(\beta)}_{kl} \in \real. \label{p4ii} \\ 
&  A_m(u;\beta)  X_k^{(\beta,N)} = A_m(u;\beta;k) X_k^{(\beta,N)} , \qquad m=1,2,\dots,N; \label{p4iii}\\  
& \text{where} \quad A_m(u;\beta;k) := \prod_{i=1}^n \frac{u+\beta^{-1}\ov{k_i}+i-n-1 + \delta(\un{k_i} \leq m) }{u+\beta^{-1}\ov{k_i}+i-n-1}. \nonumber \\ 
& \text{The $N$-tuples of rational functions in $u$:  $  A_1(u;\beta;k), \dots,  A_N(u;\beta;k) $ } \label{p4iv} \\  
& \text{are distinct for distinct $k \in \LC_n^{(1)}$.} \nonumber \\
& {\sprod{X_k^{(\beta,N)}}{X_l^{(\beta)}}}_{\beta,N} = 0 \quad \text{for} \quad k \neq l. \label{p4v}
\end{align}
\end{em}
\begin{pf}
The statement (\ref{p4i}) is a direct consequence of (\ref{p3i}), of the decomposition (\ref{eq:Yangdecomp}) and of the easily verified fact that the correspondence between the set of all pairs $( s , a )$ where $s \in \LC_n^{(N)}, $ $ a \in T_{p(s)},$ and the set of strictly decreasing sequences $\LC_n^{(1)}$ given by   
\begin{align}
& k = (k_1,k_2,\dots,k_n) \in \LC_n^{(1)}  \rightarrow  s  = (\ov{k_n},\ov{k_{n-1}},\dots,\ov{k_1}) \in \LC_n^{(N)}, \quad a = (\un{k_n},\un{k_{n-1}},\dots,\un{k_1}) \in T_{p(s)}   
\end{align}
is one-to-one.

The triangularity of $X_k^{(\beta,N)}$ expressed by  (\ref{p4ii}) is established  as follows. In view of (\ref{p3ii}) and (\ref{eq:Phi=Psi}) we have 
\begin{equation}
X_k^{(\beta,N)} = \Psi_{k}^{(\beta)} +   \sum_{l \in \LC_n^{(1)},\; \ov{l} = \ov{k},\; \un{l} < \un{k} } c_{kl}^{(\beta)} \Psi_{l}^{(\beta)} \qquad ( c_{kl}^{(\beta)} \in \real ). 
\end{equation}
Now observe that $ \ov{l} = \ov{k}, $ $\un{l} < \un{k}$ imply  $ \ov{l} = \ov{k}, $ ${l} < {k}.$ And therefore (\ref{p4ii}) follows from the triangularity of eigenvectors $\Psi_{l}^{(\beta)}$ expressed by (\ref{p1i}) and  (\ref{p1iii}). 

The equation (\ref{p4iii}) follows immediately from (\ref{p3iii}) and the explicit expression for the eigenvalues of the Dunkl operators (\ref{eq:d-eigenvalue}). 

The simplicity of the spectrum of the $A(\glN;\beta)$ expressed by the statement (\ref{p4iv}) is proved as follows. First of all, (\ref{p3iv}) shows that spectrum of $A(\glN;\beta)$ is simple on each irreducible component $F_s$ of the Yangian action $Y(\glN;\beta).$ Next, by a straightforward modification of the proof of the statement (\ref{p2v}) which is contained in Appendix, section A, one shows that the spectrum of the Quantum Determinant $A_N(u;\beta)$ separates between these irreducible components. 

Finally, the orthogonality (\ref{p4v}) is a consequence of (\ref{p4iv}) and the self-adjointness (\ref{eq:selfa}) of $A(\glN;\beta)$ relative to the scalar product ${\sprod{\cdot}{\cdot}}_{\beta,N}.$
\end{pf}

The orthogonal eigenbasis $\{ X_k^{(\beta,N)} \; | \; k \in \LC_n^{(1)} \}$ is precisely the eigenbasis of $\F$ that was described in section 1.1. To see this we simply have to observe that if we fix an integer $M$ and define for any $k \in \LC_n^{(1)} $ a sequence $\s = (\s_1,\s_2,\dots,\s_n) \in \LC_n^{(n)}$ by $
k_i = \s_i + M - i + 1 ,$ then the correspondence between this $k$ and the pair
\begin{equation}
 \left( r = \s_n \; ,\; \lambda = (\s_1 - \s_n , \s_2 - \s_n , \dots ,\s_{n-1} - \s_n, 0) \right)
\end{equation}
is bijective. In other words we may label the elements of the basis  $\{ X_k^{(\beta,N)} \; | \; k \in \LC_n^{(1)} \}$ by a pair which consists of an integer $r$ and a partition $\lambda$ with length less or equal to $n-1.$ This is the parameterization adopted in section 1.1 of the Introduction.

\section{Spin Calogero-Sutherland Model in the language of 
 symmetric polynomials} 

Starting with this section we  move into the realm of symmetric polynomials so that many objects we are about to encounter  will be parameterized by partitions. Below we summarize partition-related  notations we are going to use. Mainly we will conform to the notations of the book \cite{MacBook}.   

\subsection{Notations} \label{sec:notations}
For a partition $\lambda = (\lambda_1,\lambda_2,\dots\;)$ we will denote by $l(\lambda)$ the length i.e. the number of non-zero parts  $\lambda_1,\lambda_2,\dots $ in $\lambda.$ All partitions that appear in this paper have length less or equal to $n$ which is  the number of particles in the Spin Calogero-Sutherland Model. 

We will identify a partition $\lambda$ with its diagram defined as the set
\begin{equation}
\{ \; (i,j)\; | \;  1 \leq i \leq l(\lambda),\: 1 \leq j \leq \lambda_i \},
\end{equation}
graphically represented as collection of squares with coordinates $(i,j)$ where $i$ is increasing downward and $j$ is increasing from left to right as in the picture below which represents $\lambda = (6,4,4,3,1).$
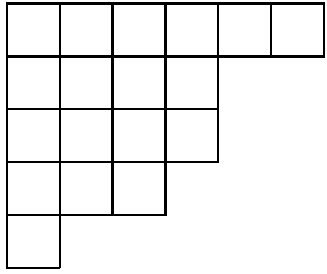
\begin{center}

\begin{picture}(120,100)
\put(0,100){\line(1,0){120}}
\put(0,80){\line(1,0){120}}
\put(0,60){\line(1,0){80}}
\put(0,40){\line(1,0){80}}
\put(0,20){\line(1,0){60}}
\put(0,0){\line(1,0){20}}
\put(0,0){\line(0,1){100}}
\put(20,0){\line(0,1){100}}
\put(40,100){\line(0,-1){80}}
\put(60,100){\line(0,-1){80}}
\put(80,100){\line(0,-1){60}}
\put(100,100){\line(0,-1){20}}
\put(120,100){\line(0,-1){20}}
\end{picture}  \\
\end{center}

For a square $s \in \lambda$  arm-length $ a_{\lambda}(s)$, leg-length $ l_{\lambda}(s)$, arm-colength $ a'(s)$ and leg-colength  $ l'(s)$ are defined as the number of squares in the diagram of $\lambda$ to the east, south, west and north from $s$ respectively. Also for a square $s$ we define
\begin{itemize} 
\item Content: $ c(s) =  a'(s) - l'(s).$
\item Hook-length: $h_{\lambda}(s) =  a_{\lambda}(s) + l_{\lambda}(s) + 1.$
\end{itemize}
And for a real number $\gamma$ their refinements:
\begin{gather}
c(s;\gamma) =  a'(s) - \gamma l'(s), \label{eq:rcontent}\\
h_{\lambda}^*(s;\gamma) =  a_{\lambda}(s) + \gamma l_{\lambda}(s) + 1, \quad 
h^{\lambda}_*(s;\gamma) =  a_{\lambda}(s) + \gamma l_{\lambda}(s) + \gamma. \label{eq:rhook}
\end{gather}

For a positive integer $N$ which as before has the meaning of the number of spin degrees of freedom in the Spin Calogero-Sutherland Model we define the following two subsets of $\lambda$:
\begin{align}
& C_N(\lambda) = \{ s \in \lambda \;|\; c(s) = 0\bmod N \}, \\  
& H_N(\lambda) = \{ s \in \lambda \;|\; h_{\lambda}(s) = 0\bmod N \}.
\end{align}
And for any subset of squares  $S \subset \lambda$ we denote by $| S |$ the number of squares in $S.$

\subsection{An isomorphism between the space of states of the Spin Calogero-Sutherland Model and the space of symmetric Laurent polynomials} \label{sec:isomorphism}
Here we discuss main technical points  of this paper -- the definition and properties of the isomorphism between the space of states of the Spin Calogero-Sutherland Model and the space of symmetric Laurent polynomials.

To avoid having to deal with questions of convergence of various integrals, we assume from now and until the end of this paper that  $\beta$ is a non-negative integer number. With this assumption it will be possible to keep our discussion completely algebraic. 

Let $L_n =\cx$ be the $\cplx$-algebra of Laurent polynomials in variables $x_1,\dots,x_n.$ If $f$ $=$ $f(x_1,\dots,x_n)$ $\in$ $L_n$ let $ \overline{f(x_1,\dots,x_n)}$ be the Laurent polynomial with complex conjugated coefficients, let $f^* $ $ =$ $  \overline{f(x_1^{-1},\dots,x_n^{-1})},$ and let $[f]_1$ denote the constant term in $f$.  

Let $A_n^{\pm}$ be the subspace of skew-symmetric Laurent polynomials in $L_n$. For $l =(l_1,\dots,l_n)$ $\in$ $\zint^n$ define the antisymmetric monomial $\hat{a}_l$ analogous to the wedge vector (\ref{eq:antisymm}) as follows: 
\begin{equation}
\hat{a}_l = x^{l_1}\wedge x^{l_2} \wedge \cdots \wedge x^{l_n} = \sum_{w \in S_n}\sgn(w)  x^{l_{w(1)}}_1 x^{l_{w(2)}}_2  \cdots  x^{l_{w(n)}}_n.
\end{equation}
Then $\{\hat{a}_l\:|\: l\in \LC_n^{(1)} \}$ is a basis of  $A_n^{\pm}.$

Now let us introduce
\begin{equation}
 \tilde{\Delta}(\beta,N) := \prod_{1\leq i \neq j \leq n} ( 1 - x_i^N x_j^{-N})^{\beta}.
\end{equation}
Since $\beta$ is a non-negative integer, $\tilde{\Delta}(\beta,N)$ is a symmetric Laurent polynomial.
We define a scalar product ${\sproddd{\cdot}{\cdot}}_{\beta,N}$ on $L_n$ by setting for $f,g$ $\in$ $L_n$:
\begin{equation}
 {\sproddd{f}{g}}_{\beta,N} = \frac{1}{n!}[f^*\tilde{\Delta}(\beta,N)g]_1. \label{eq:scAL}
\end{equation}
Let us use  $ {\sproddd{f}{g}}_0 $ as a short-hand notation for $ {\sproddd{f}{g}}_{0,N}, $ then we have
\begin{equation}
{\sproddd{f}{g}}_{\beta,N} = {\sproddd{f}{\tilde{\Delta}(\beta,N) g}}_0, \label{eq:e1}
\end{equation}
where we regard the $\tilde{\Delta}(\beta,N)$ as a multiplication operator on $L_n.$

When $\beta$ is a non-negative integer, the definition of the scalar product $ {\sprod{\cdot}{\cdot}}_{\beta}' $ (\ref{eq:spc}) may be formulated as follows: for $f,g$ $\in$ $\cz$:  
\begin{equation}
{\sprod{f}{g}}_{\beta}' = \frac{1}{n!}[f^*{\Delta}(\beta)g]_1,
\end{equation}
where $ {\Delta}(\beta) =\prod_{1\leq i \neq j \leq n} ( 1 - z_i z_j^{-1})^{\beta} $ is now a symmetric Laurent polynomial. Then for the scalar product  ${\sprod{\cdot}{\cdot}}_{\beta,N}$ on the space $\F$ we have with $f,g$ $\in$ $\F$:
\begin{equation}
 {\sprod{f}{g}}_{\beta,N} = {\sprod{f}{{\Delta}(\beta)g}}_{0,N}, \label{eq:e2}
\end{equation}
where we consider the $ {\Delta}(\beta)$ as a multiplicaltion operator on the space $\F.$ Note that in  the last formula we have ${\Delta}(\beta){g} \in \F$ since $ {\Delta}(\beta)$ is a symmetric Laurent polynomial. 

Let now $\omega'$ $:$ $\F \rightarrow A_n^{\pm}$ be a $\cplx$-map defined for $k \in \zint^n$ by $ \omega'(\hat{u}_k) = \hat{a}_k.$ Clearly this map is an isomorphism of linear spaces. Moreover for any $f,g \in \F$ we have: 
\begin{equation}
 {\sprod{f}{g}}_{0,N} = {\sproddd{\omega'(f)}{\omega'(g)}}_0, \label{eq:strange1}  \label{eq:e3}
\end{equation}
for if $k,l \in \LC_n^{(1)}$ then 
\begin{equation}
{\sprod{\hat{u}_k}{\hat{u}_l}}_{0,N} = \delta_{kl} =  {\sproddd{\hat{a}_k}{\hat{a}_l}}_0
\end{equation}
which shows (\ref{eq:strange1}) since $\{ \hat{u}_k \: | \: k \in \LC_n^{(1)} \}$ and $\{ \hat{a}_k \: | \: k \in \LC_n^{(1)} \}$ are bases of the spaces $\F$ and  $A_n^{\pm}$ respectively.

Now we are ready to formulate one of the crucial technical points of this paper, which is the following relation between the scalar product ${\sprod{\cdot}{\cdot}}_{\beta,N}$ ( sec. \ref{sec:scalar}) on the space of states $\F$  of the gauge-transformed  Spin Calogero-Sutherland Model and the scalar product (\ref{eq:scAL}) on the space of skew-symmetric Laurent polynomials $A_n^{\pm}$:
\begin{gather}
\text{ {\em For any $f,g$ $\in$ $\F$ we have }} \quad  {\sprod{f}{g}}_{\beta,N} = {\sproddd{\omega'(f)}{\omega'(g)}}_{\beta,N}.\label{eq:s=s}
\end{gather}
\begin{pf}
Let $P = P(z_1,\dots,z_n) \in \cz$ be a symmetric Laurent polynomial. Then for any $f \in \F$ the following relation holds 
\begin{equation}
\omega'( P(z_1,\dots,z_n) f ) = P(x_1^{-N},\dots,x_n^{-N})\omega'( f ). \label{eq:pf1}
\end{equation}
Indeed, the algebra $(\cz)^{S_n}$ of symmetric Laurent polynomials in variables $z_1,\dots,z_n$ is generated by the power-sums:
\begin{equation}
p_r = p_r(z_1,\dots,z_n) = \sum_{i=1}^n z_i^r, \qquad (r=0,1,2,\dots\;) \label{eq:pws}
\end{equation}
and the element 
\begin{equation}
z_1^{-1}z_2^{-1}\cdots z_n^{-1}. \label{eq:shift} 
\end{equation}
So it is enough to show (\ref{eq:pf1}) when $ P$ equals to one of the Laurent polynomials (\ref{eq:pws}), ( \ref{eq:shift}). Also, since $\hat{u}_k,$ $ k \in \zint^n$ span $\F$ we may assume that in (\ref{eq:pf1}) $ f = \hat{u}_k.$ Calculating:
\begin{align}
& \omega'(p_r \hat{u}_k) = \omega'\left( \sum_{i=1}^n u_{k_1}\wedge \cdots \wedge u_{k_i - Nr}\wedge \cdots \wedge u_{k_n} \right) = \\ 
 & = \sum_{i=1}^n x^{k_1}\wedge \cdots \wedge x^{k_i - Nr}\wedge \cdots \wedge x^{k_n} = p_r(x_1^{-N},\dots,x_n^{-N}) \hat{a}_k; \nonumber \\ 
&\text{and} \quad  \omega'(z_1^{-1}z_2^{-1}\cdots z_n^{-1} \hat{u}_k) = \omega'(  u_{k_1+N}\wedge  u_{k_2+N}\wedge \cdots \wedge u_{k_n + N} ) = \\  
& = x^{k_1+N}\wedge  x^{k_2+N}\wedge \cdots \wedge x^{k_n + N} = x_1^N x_2^N \cdots x_n^N \hat{a}_k,  \nonumber 
\end{align}
we confirm (\ref{eq:pf1}).

Now since $\Delta(\beta)$ is a symmetric Laurent polynomial, we have 
for any $f,g$ $\in$ $\F$: 
\begin{align}
& {\sproddd{\omega'(f)}{\omega'(g)}}_{\beta,N} = \text{ ( by \ref{eq:e1})} = {\sproddd{\omega'(f)}{\tilde{\Delta}(\beta,N) \omega'(g)}}_0 = \text{ (by \ref{eq:pf1})} = \nonumber \\ 
& = {\sproddd{\omega'(f)}{\omega'({\Delta}(\beta) g)}}_0 =\text{ (by \ref{eq:e3})} =  {\sprod{f}{{\Delta}(\beta) g}}_{0,N} =\text{ (by \ref{eq:e2})} = {\sprod{f}{g}}_{\beta,N}. \nonumber 
\end{align} 
\end{pf}

Now let $\Lambda_n^{\pm} = (\cx)^{S_n}$ be the linear space of symmetric Laurent polynomials in variables $x_1,\dots,x_n$ and introduce on $\Lambda_n^{\pm}$ a scalar product ${\sprodd{\cdot}{\cdot}}_{\beta,N}$ as follows: for $f,g$ $\in$  $\Lambda_n^{\pm}$ set
\begin{gather}
{\sprodd{f}{g}}_{\beta,N} = \frac{1}{n!}[f^* \Delta(\beta,N) g]_1, \\
\text{where}\quad \Delta(\beta,N) = \tilde{\Delta}(\beta,N)\prod_{1\leq i \neq j \leq n}(1 - x_i x_j^{-1}) = \prod_{1\leq i \neq j \leq n}(1 - x_i^N x_j^{-N})^{\beta}(1 - x_i x_j^{-1}). \label{eq:spbN}
\end{gather}

Let $K\in \zint$ and $\omega'_{K}$ $:$ $A_n^{\pm}$ $\rightarrow$ $\Lambda_n^{\pm}$ be an isomorphism of linear spaces defined for $f\in A_n^{\pm}$ by:  
\begin{equation}
\omega'_{K}(f) = x_1^{-K} x_2^{-K}\cdots  x_n^{-K} f /\hat{a}_{\delta},
\end{equation}
where $\hat{a}_{\delta} = \prod_{1\leq i < j \leq n}( x_i - x_j) $ is the Vandermonde determinant. Then for any $f,g \in A_n^{\pm} $ we have: 
\begin{equation}
{\sprodd{\omega'_{K}(f)}{\omega'_{K}(g)}}_{\beta,N} = {\sproddd{f}{g}}_{\beta,N}. \label{eq:s=s2}
\end{equation}

Now set $\Omega_K = \omega'_{K}\omega'.$ The map $\Omega_K$ for any integer $K$  is an isomorphism of linear spaces $\F$ and  $\Lambda_n^{\pm}$. Moreover combining (\ref{eq:s=s}) and (\ref{eq:s=s2}) we arrive at the following conclusion:
\begin{equation}
\text{{\em For any $f,g \in \F$ we have }} \quad  {\sprodd{\Omega_{K}(f)}{\Omega_{K}(g)}}_{\beta,N} = {\sprod{f}{g}}_{\beta,N}. \label{eq:s=s3} 
\end{equation}
That is $\F$ with the scalar product ${\sprod{\cdot}{\cdot}}_{\beta,N}$ and the space of symmetric Laurent polynomials $\Lambda_n^{\pm}$ with the scalar product  ${\sprodd{\cdot}{\cdot}}_{\beta,N}$ are isomorphic as Hilbert spaces with the isomorphism defined by the map $ \Omega_{K}$ where $K$ is an arbitrary integer.

\subsection{Basis of the space of symmetric Laurent polynomials defined by the $A(\glN;\beta)$-eigenbasis of the space of states $\F$} \label{sec:basis}
Fix $M\in \zint$ and for any $k = (k_1,\dots,k_n)$ $\in $ $\LC_n^{(1)}$ define $ \s = (\s_1,\dots,\s_n)$ $\in $ $\LC_n^{(n)}$  by $ k_i = \s_i + M - i + 1.$ Then with the sequence $\delta = (n-1,\dots,1,0)$ we have 
\begin{equation}
\Omega_{M-n+1}(\hat{u}_k) = (x_1\cdots x_n)^{-(M-n+1)} \hat{a}_k/\hat{a}_{\delta} = \hat{a}_{\s+\delta}/\hat{a}_{\delta} \equiv s_{\s}.
\end{equation}
In particular, when $\s_n \geq 0 $, the $s_{\s}$ is just the Schur polynomial labelled by the partition $\s.$ Note that $ \Omega_{M-n+1}(\vac(M)) = 1.$

Denote now $ \Omega_{M-n+1}(X_k^{(\beta,N)})$ by $ Y_{\s}^{(\beta,N)} $.
Then from (\ref{p3i})  - (\ref{p3iv}) we obtain:
\begin{em}
\begin{align}
& \text{$\{Y_{\s}^{(\beta,N)} \:| \: \s \in \LC_n^{(n)}\} $ is a basis of $\Lambda_n^{\pm}$ such that:} \label{eq:Ybas}\\ 
& Y_{\s}^{(\beta,N)} = s_{\s} + \sum_{\tau < \s} v_{\s\tau}^{(\beta)} s_{\tau}, \quad v_{\s\tau}^{(\beta)}  \in \real; \label{eq:tri}\\ 
& {\sprodd{Y_{\s}^{(\beta,N)}}{Y_{\tau}^{(\beta,N)}}}_{\beta,N} = 0 \quad \text{if} \quad \s \neq \tau. \label{eq:ort}
\end{align}
\end{em}
Note that the basis $\{Y_{\s}^{(\beta,N)} \:| \: \s \in \LC_n^{(n)}\} $ with the properties (\ref{eq:tri}) and (\ref{eq:ort})  is unique. This follows from the standard argument \cite{MacBook} based on the Gram-Schmidt orthogonalization of the basis of the Laurent polynomials $s_{\s}.$ Note also that $\{ Y_{\lambda}^{(\beta)} \} $ where $\lambda$ runs through all partitions of length less or equal to $n$, is a basis of the space of symmetric polynomials $\Lambda_n = (\cplx[x_1,\dots,x_n])^{S_n}$.

\subsection{$\glN$-Jack Polynomials}  \label{sec:jack}
Let $q$ and $t$ be parameters and let $\Lambda_n^{q,t} = (\cplx(q,t)[x_1,\dots,x_n])^{S_n}$ be the $\cplx(q,t)$-algebra of symmetric polynomials in variables $x_1,\dots,x_n$. For $f=f(x_1,\dots,x_n),$ $g=g(x_1,\dots,x_n)$ $\in$ $\Lambda_n^{q,t}$ a scalar product is defined as follows \cite{MacBook}:  
\begin{equation}
{\sprodd{f}{g}}_{q,t} = \frac{1}{n!}\prod_{j=1}^n \int \frac{dw_j}{2\pi i w_j} \prod_{1\leq i \neq j \leq n}\frac{ (w_i w_j^{-1} ; q)_{\infty}}{ (t w_i w_j^{-1} ; q)_{\infty}} \overline{f(w_1,\dots,w_n)}g(w_1,\dots,w_n), \label{eq:MPsp}
\end{equation}
where the integration in each of the variables $w_j$ is taken along the unit circle in the complex plane, and $ (x ; q )_{\infty} = \prod_{r = 0}^{\infty} ( 1 - x q^r).$  Then for each partition $\lambda$ of length less or equal to $n$ the Macdonald Polynomial $P_{\lambda}(q,t) =P_{\lambda}(x_1,\dots,x_n ; q,t) $ \cite{MacBook} is uniquely defined as the element of $\Lambda_n^{q,t}$ such that:   
\begin{em}\begin{gather}
 P_{\lambda}(q,t) = m_{\lambda} + \sum_{\mu < \lambda} u_{\lambda\mu}(q,t) m_{\mu}, \label{eq:MP=m+} \\
 \text{ where $m_{\lambda} = m_{\lambda}(x_1,\dots,x_n)$ is the monomial symmetric polynomial and $ u_{\lambda\mu}(q,t)  \in \cplx(q,t);$} \nonumber \\ 
 {\sprodd{P_{\lambda}(q,t)}{P_{\mu}(q,t)}}_{q,t} = 0 \quad \text{if} \quad \lambda \neq \mu.  \label{eq:MPort}
\end{gather}\end{em}
Let now $\gamma $ be a positive real number, let $\o_N = \exp(\frac{2\pi i}{N})$ and consider the limit:  
\begin{equation}
 q = \o_N p, \quad t = \o_N p^{\gamma}, \qquad p \rightarrow 1. \label{eq:limit}
\end{equation}
In this limit the Macdonald Polynomial $P_{\lambda}(q,t)$ degenerates into  a symmetric polynomial which we will denote by  $P_{\lambda}^{(\gamma,N)}$ and call the {\em $\glN$-Jack Polynomial}. In particular when $N=1$ the $\gl_1$-Jack Polynomial is nothing but the usual Jack Polynomial: $P_{\lambda}^{(\gamma,1)}$$= $  $P_{\lambda}^{(\gamma^{-1})}$ in the notation of \cite{MacBook}.

First, let us verify that the polynomial $P_{\lambda}^{(\gamma,N)}$ is well-defined. To do this we will take the limit (\ref{eq:limit}) in the coefficients $u_{\lambda\mu}(q,t)$ of the expansion (\ref{eq:MP=m+}). 

For a partition $\lambda$ a tableau $T$ of shape $\lambda$ \cite{MacBook} is a sequence of partitions:
\begin{equation}
\emptyset = \lambda^{(0)} \subset   \lambda^{(1)} \subset  \dots \subset  \lambda^{(r)} = \lambda  
\end{equation}
such that each skew diagram $\theta^{(i)} = \lambda^{(i)} - \lambda^{(i-1)}$ $(1 \leq i \leq r)$ is a horizontal strip. The sequence $ ( |\theta^{(1)}|,\dots,|\theta^{(r)}|)$ is called the weight of $T.$  The coefficient of $m_{\mu}$ in the expansion of $P_{\lambda}(q,t)$  is \cite{MacBook}:
\begin{equation} 
u_{\lambda\mu}(q,t) = \sum_{T} \psi_{T}(q,t) 
\end{equation}
summed over tableaus of shape $\lambda$ and weight $\mu.$ To describe the $\psi_{T}(q,t)$, for partitions $\lambda$ and $\mu$ such that $\mu \subset \lambda$ let $C_{\lambda/\mu}$ (resp. $R_{\lambda/\mu}$) denote the union of the columns (resp. rows) that intersect $\lambda - \mu.$ Then:  
\begin{equation}
\psi_{T}(q,t) = \prod_{i=1}^r \psi_{\lambda^{(i)}/\lambda^{(i-1)}}(q,t),
\end{equation}
where for $\mu \subset \lambda$: 
\begin{align}
& \psi_{\lambda/\mu}(q,t) = \prod_{s \in R_{\lambda/\mu} - C_{\lambda/\mu}} \frac{b_{\mu}(s;q,t)}{b_{\lambda}(s;q,t)}, \\ 
& \text{and} \quad b_{\lambda}(s;q,t) = \begin{cases} \frac{1 - q^{a_{\lambda}(s)}t^{l_{\lambda}(s) + 1}}{1 - q^{a_{\lambda}(s)+ 1}t^{l_{\lambda}(s)}} &  \text{if $ s\in \lambda,$ } \\  1 & \text{otherwise.} \end{cases}
\end{align}
Now taking the limit (\ref{eq:limit}) in $b_{\lambda}(s;q,t)$ we obtain:
\begin{equation}
b_{\lambda}^{(\gamma,N)} (s) = \lim_{p\rightarrow 1}b_{\lambda}(s; \o_N p,\o_N p^{\gamma}) = \begin{cases} \frac{ a_{\lambda}(s) + \gamma l_{\lambda}(s) + \gamma }{ a_{\lambda}(s) + \gamma l_{\lambda}(s) + 1 }  & \text{ if $ s \in \lambda,$ $ h_{\lambda}(s) = 0\bmod N;$} \\ 1 & \text{otherwise.} \end{cases}  \label{eq:b}
\end{equation} 
Thus the coefficient $u_{\lambda\mu}(q,t)$ in (\ref{eq:MP=m+}) has a well-defined limit $u_{\lambda\mu}^{(\gamma,N)}$ such that $u_{\lambda\mu}^{(\gamma,N)}$ $ >0 $ for $\gamma > 0$, and 
\begin{equation}
 P_{\lambda}^{(\gamma,N)} =  m_{\lambda} + \sum_{\mu < \lambda} u_{\lambda\mu}^{(\gamma,N)} m_{\mu}. \label{eq:P=m+}
\end{equation}
Sometimes it is more convenient to use the expansion of the polynomial $P_{\lambda}^{(\gamma,N)}$ in the basis of Schur polynomials $ s_{\lambda}$ which is also unitriangular in the natural ordering of partitions:
\begin{equation}
 P_{\lambda}^{(\gamma,N)} =  s_{\lambda} + \sum_{\mu < \lambda} v_{\lambda\mu}^{(\gamma,N)} s_{\mu},  \qquad   v_{\lambda\mu}^{(\gamma,N)} \in \real,  \label{eq:P=s+}
\end{equation}
since the bases $\{ m_{\lambda} \} $ and $ \{ s_{\lambda} \}$ are related by a unitriangular transformation \cite{MacBook}. Note also, that $ P_{\lambda}(q,q) = s_{\lambda} $ \cite{MacBook}  gives $ P_{\lambda}^{ (1,N)} = s_{\lambda}.$

Consider next the behaviour of the scalar product (\ref{eq:MPsp}) in the limit (\ref{eq:limit}). To avoid the question of convergence of the integral in (\ref{eq:MPsp}) in this limit, we now assume that $ t = q^k $ where $k$ is a non-negative integer. Then for any $f,g$ $\in$ $\Lambda_n^{\pm}$ we have \cite{MacBook}:   
\begin{align}
& {\sprodd{ f}{g}}_{q,t} = \frac{1}{n!} [ f^* \Delta(q,t)g ]_1 ,  \\
& \text{where} \quad \Delta(q,t) = \prod_{1\leq i \neq j \leq n} \frac{ (x_i x_j^{-1} ; q)_{\infty}}{ (t x_i x_j^{-1} ; q)_{\infty}} =  \prod_{1\leq i \neq j \leq n} \prod_{r=0}^{k-1} ( 1 - q^r x_i x_j^{-1}). \label{eq:MPsp1} 
\end{align}
Let now $k=\gamma$ and let $\gamma = N\beta + 1$ where $\beta$ is a non-negative integer. Then taking the limit (\ref{eq:limit}) in (\ref{eq:MPsp1}) we obtain: 
\begin{equation}
\Delta(\beta,N) = \lim_{p\rightarrow 1}\Delta(\o_N p, \o_N p^{\gamma} ) =  \prod_{1\leq i \neq j \leq n} \prod_{r=0}^{N\beta}(1 - \o_N^r x_i x_j^{-1}) = \prod_{1\leq i \neq j \leq n} (1 -  x_i^N x_j^{-N})^{\beta}  (1 -  x_i x_j^{-1}).
\end{equation}
This is exactly the weight function of the scalar product ${\sprodd{\cdot}{\cdot}}_{\beta,N}$ which was introduced in (\ref{eq:spbN}). Thus for $ \gamma = N\beta + 1 $ the scalar product for Macdonald Polynomials ${\sprodd{\cdot}{\cdot}}_{q,t}$ degenerates in the limit (\ref{eq:limit}) into the scalar product ${\sprodd{\cdot}{\cdot}}_{\beta,N}$ defined in (\ref{eq:spbN}) so that from (\ref{eq:MPort}) we get   
\begin{equation}
{\sprodd{P_{\lambda}^{(N\beta+1,N)}}{P_{\mu}^{(N\beta+1,N)}   }}_{\beta,N} = 0, \quad \text{if $ \lambda \neq \mu $.} \label{eq:Port}
\end{equation}

It is now clear that for the $\glN$-Jack Polynomials we may establish analogues of all properties known for Macdonald Polynomials just by taking the limit (\ref{eq:limit}). We will not go here into discussion of the Cauchy formulas, Duality etc. \cite{MacBook} all of which are straightforward to derive for the  $\glN$-Jack Polynomials. In this paper we will restrict ourselves to computing the normalization and the expansion of power-sums in terms of  $P_{\lambda}^{(\gamma,N)}$ because these are relevant to computation of Dynamical Spin-Density and Density Correlation Functions in the Spin Calogero-Sutherland Model.

First of all taking the limit (\ref{eq:limit}) in the norm formula for the Macdonald Polynomial \cite{MacBook}:
\begin{gather}
{\sprodd{P_{\lambda}(q,t)}{P_{\lambda}(q,t)}}_{q,t} = c_n(q,t)\prod_{ s\in \lambda} \frac{1 - q^{a'(s)}t^{n - l'(s)}}{1 - q^{a'(s)+ 1}t^{n - l'(s) -1 }} \prod_{ s\in \lambda} \frac{1 - q^{a_{\lambda}(s)+1}t^{l_{\lambda}(s)}}{1 - q^{a_{\lambda}(s)}t^{l_{\lambda}(s)+1} }, \nonumber \\ 
c_n(q,t) = {\sprodd{ 1 }{ 1 }}_{q,t} = \prod_{1 \leq i < j \leq n}\frac{ (t^{j-i} ; q)_{\infty}(q t^{j-i} ; q)_{\infty}}{ (t^{j-i+1} ; q)_{\infty}(q t^{j-i-1} ; q)_{\infty}}, \nonumber 
\end{gather}
we obtain the norm formula for the $\glN$-Jack Polynomial:
\begin{multline}
{\sprodd{P_{\lambda}^{(\gamma,N)}}{P_{\lambda}^{(\gamma,N)}}}_{\beta,N} = \\ = c_n^{(\gamma,N)}\prod\begin{Sb} s\in \lambda \\ c(s) = n\bmod N \end{Sb} \frac{a'(s) + \gamma ( n - l'(s))}{a'(s)+ 1 + \gamma ( n - l'(s) - 1 )} \prod\begin{Sb} s\in \lambda \\ h_{\lambda}(s) = 0\bmod N\end{Sb} \frac{a_{\lambda}(s) + \gamma l_{\lambda}(s) + 1 }{a_{\lambda}(s) + \gamma l_{\lambda}(s)+ \gamma  }.  \label{eq:norm}
\end{multline}
In the last formula we have set $\gamma = N\beta + 1$ and 
\begin{gather}
 c_n^{(\gamma,N)} = {\sprodd{1}{1}}_{\beta,N} = \prod_{1\leq i < j \leq n} C^{(\gamma, N)}(j-i)\qquad  \text{where} \\ 
 C^{(\gamma, N)}(k) = \begin{cases} \frac{\Gamma\left(\frac{ \gamma k - \gamma 
+ 1 - a }{N} + 1\right) \Gamma\left(\frac{ \gamma k + \gamma - 1 - a }{N} + 1\right) }{ \Gamma^2\left(\frac{ \gamma k - a }{N} + 1\right)} & \text{when $ k = a \bmod N$ and $ a =1,\dots,N-1;$ } \\ 
\frac{\Gamma\left(\frac{ \gamma k - \gamma + 1 }{N} + 1\right) \Gamma\left(\frac{ \gamma k + \gamma - 1 }{N} + 1\right) }{ \Gamma\left(\frac{ \gamma k }{N} + 1\right)\Gamma\left(\frac{ \gamma k }{N} \right)} & \text{when $ k =0\bmod N.$}
\end{cases} \end{gather}
In the paper \cite{TU} the norm formulas for eigenvectors of the commutative family $A(\glN;\beta)$ were computed  by other, rather cumbersome method. For the case $N=2$ we have verified that the formulas in \cite{TU} give exactly the same result as the formula above.

Consider now the formula for the expansion of the power-sums in the basis of Macdonald Polynomials \cite{Konno}: 
\begin{equation}
p_m = \sum_{i=1}^n x_i^m = (1-q^m)\sum\begin{Sb} \lambda \\ |\lambda| = m \end{Sb} \frac{ \prod_{ s \in \lambda \setminus  (1,1) } t^{l_{\lambda}'(s)} - q^{a_{\lambda}'(s)} }{ \prod\begin{Sb} s \in \lambda  \end{Sb} 1 -  t^{l_{\lambda}(s)}  q^{a_{\lambda}(s) + 1 } } P_{\lambda}(q,t), \quad (m=0,1,2,\dots \;).  \label{eq:p=MP}
\end{equation}
For any partition $\lambda$ let us define the subsets: 
\begin{align}
& C_N(\lambda) = \{ s \in \lambda \;|\; c(s) = 0\bmod N \}, \\  
& H_N(\lambda) = \{ s \in \lambda \;|\; h_{\lambda}(s) = 0\bmod N \}.
\end{align}
Then taking the limit (\ref{eq:limit}) in the formula (\ref{eq:p=MP}) we obtain the formula for expansion of the power-sums in the basis of $\glN$-Jack Polynomials:
\begin{equation}
p_m = \sum\begin{Sb} \lambda \\ |\lambda | = m \end{Sb} \chi_{\lambda}^{(\gamma,N)} P_{\lambda}^{(\gamma,N)}, \label{eq:p=P+}
\end{equation}
where the coefficients $ \chi_{\lambda}^{(\gamma,N)} $ are as follows:
\begin{align}
\chi_{\lambda}^{(\gamma,N)}& =  
|\lambda | \o_N^{n(\lambda)} \frac{ \prod_{ s \in \lambda \setminus C_N(\lambda)} (1 - \o_N^{c(s)})}{ \prod_{ s \in \lambda \setminus H_N(\lambda)} (1 - \o_N^{h_{\lambda}(s)})} \frac{ \prod_{ s \in C_N(\lambda)\setminus (1,1) } ( a'(s) - \gamma l'(s))} { \prod_{ s \in H_N(\lambda) } (a_{\lambda}(s) + \gamma l_{\lambda}(s) + 1)}  \label{eq:psexp1}\\
& \text{ when $ |\lambda | = 0\bmod N, \quad |C_N(\lambda)| = |H_N(\lambda)|;$}  \nonumber \\ 
\chi_{\lambda}^{(\gamma,N)}& =    0 \label{eq:psexp2} \\  
& \text{ when $ |\lambda | = 0\bmod N, \quad   |C_N(\lambda)| > |H_N(\lambda)|;$}\nonumber  \\ 
\chi_{\lambda}^{(\gamma,N)}& =  (1 - \o_N^{|\lambda|})\o_N^{n(\lambda)} \frac{ \prod_{ s \in \lambda \setminus C_N(\lambda)} (1 - \o_N^{c(s)})}{ \prod_{ s \in \lambda \setminus H_N(\lambda)}(1 - \o_N^{h_{\lambda}(s)})} \frac{ \prod_{ s \in C_N(\lambda)\setminus (1,1) } (a'(s) - \gamma l'(s))}{ \prod_{ s \in H_N(\lambda) } (a_{\lambda}(s) + \gamma l_{\lambda}(s) + 1)} \label{eq:psexp3}\\
& \text{ when $ |\lambda | \neq  0\bmod N,\quad  |C_N(\lambda)| = |H_N(\lambda)|+1;$} \nonumber  \\ 
\chi_{\lambda}^{(\gamma,N)} & =   0  \label{eq:psexp4}\\ 
& \text{ when $ |\lambda | \neq  0\bmod N,\quad |C_N(\lambda)| > |H_N(\lambda)|+1.$} \nonumber 
\end{align}
Here $n(\lambda) = \sum (i-1)\lambda_i .$ 

Note that for any partition $\lambda$ such that $|\lambda| = 0\bmod N$  we have $|C_N(\lambda)| \geq |H_N(\lambda)|.$  And for any partition $\lambda$ such that $|\lambda| \neq 0\bmod N$  we have $|C_N(\lambda)| \geq |H_N(\lambda)|+1.$ 

\subsection{Identification of  $\glN$-Jack polynomials and polynomials $Y_{\lambda}^{(\beta,N)}$} \label{sec:identification}
Here we identify the symmetric Laurent polynomials $ Y_{\s}^{(\beta,N)}$ (\ref{eq:Ybas}) with the  $\glN$-Jack polynomials multiplied by a certain power of the Laurent polynomial $ x_1 x_2 \cdots x_n .$ This is the final step in our construction, and it  gives us the proof of the main result of this paper.

For $\s \in \LC_n^{(n)}$ $=$ $ \{ \s = (\s_1,\dots,\s_n) \in \zint^n \: | \: \s_1 \geq \s_2 \geq \cdots \geq \s_n \}$ let us define a symmetric Laurent polynomial $P_{\sigma}(q,t)$ by the relation:
\begin{equation}
P_{\sigma}(q,t) = (x_1 x_2 \cdots x_N)^{\s_n} P_{(\sigma - \s_n I_n) }(q,t).
\end{equation}
Here $(\sigma - \s_n I_n) = (\s_1 -\s_n , \s_2 -\s_n , \dots ,\s_{n-1}-\s_n,0)$ is a partition and $P_{(\sigma - \s_n I_n) }(q,t)$ is the corresponding Macdonald Polynomial. If in $\s$ all $\s_i$ are non-negative, then $\s$ is identified with a partition, say $\lambda$,  of length less or equal to $n$ and $P_{\sigma}(q,t) = P_{\lambda}(q,t)$ since \cite{MacBook} 
\begin{equation}
x_1 x_2 \cdots x_n  P_{\lambda}(q,t) = P_{(\lambda + I_n)}(q,t)
\end{equation}
for any partition $\lambda$ such that $l(\lambda) \leq n.$  

Similarly for any $\s \in \LC_n^{(n)}$ we define   
\begin{align}
& P_{\sigma}^{(\gamma,N)} = (x_1 x_2 \cdots x_N)^{\s_n} P_{(\sigma - \s_n I_n) }^{(\gamma,N)},  \label{eq:LPP} \\
& s_{\s} = (x_1 x_2 \cdots x_N)^{\s_n}s_{(\sigma - \s_n I_n)} = \hat{a}_{\s +\delta}/\hat{a}_{\delta}. 
\end{align}
Now from (\ref{eq:P=s+}, \ref{eq:Port}) we obtain:
\begin{em}\begin{align}
& P_{\sigma}^{(\gamma,N)} = s_{\s} + \sum_{\tau < \s} v_{\s\tau}^{(\gamma,N)}  s_{\tau}, \qquad ( v_{\s\tau}^{(\gamma,N)} \in \real ), \\
& {\sprodd {P_{\sigma}^{(N\beta + 1,N)}}{P_{\tau}^{(N\beta + 1,N)}}}_{\beta,N} = 0 \quad \text{if $ \s \neq \tau .$}   
\end{align} \end{em}
Comparing this with (\ref{eq:Ybas} - \ref{eq:ort}) and taking into account the uniqueness of  Laurent polynomials staisfying (\ref{eq:Ybas} - \ref{eq:ort}), we establish the main result of this paper which is equivalent to the statement given in section 1.1 of the Introduction.

{\large {\bf Theorem }} 
\begin{em}
For any $\beta \in \nat$ and $M \in \zint$ the Laurent polynomials (\ref{eq:Ybas}) obtained from the eigenvectors of the  commutative family $A(\glN;\beta)$ under the action of the isomorphism $\Omega_{M-n+1}$ $:$ $\F$ $\rightarrow$ $\Lambda_n^{\pm}$ are identical with the Laurent polynomials (\ref{eq:LPP}) where $\gamma = N \beta + 1$,i.e:  
\begin{equation}
Y_{\s}^{(\beta,N)} = P_{\sigma}^{(N\beta + 1,N)}, \qquad \s \in \LC_n^{(n)}. \label{eq:Y=P}
\end{equation}
In particular for any partition $\lambda$ of length less or equal to $n$ the symmetric polynomial $Y_{\lambda}^{(\beta,N)}$ is identical with the $\glN$-Jack Polynomial $ P_{\lambda}^{(N\beta + 1,N)}.$
\end{em}

To end this section let us now note, that the Laurent polynomials $P_{\s}(q,t)$ are eigenvectors of the Macdonald Operator \cite{MacBook}:
\begin{align}
& D_1(q,t)  = \sum_{i=1}^n \prod_{j \neq i} \frac{ t x_i - x_j}{x_i - x_j} q^{x_i\frac{\p}{\p x_i}}, \\
& \text{so that} \quad D_1(q,t)P_{\s}(q,t) = E_{\s}(q,t) P_{\s}(q,t), \\
& \text{where} \quad  E_{\s}(q,t) = \sum_{i=1}^n q^{\s_i} t^{n-i}.
\end{align}
And let the map $\chi:$ $\Lambda_n^{\pm}$ $\rightarrow$  $\Lambda_n^{\pm}$ be defined on any Laurent polynomial $f(x_1,\dots,x_n)$ by $ \chi f(x_1,\dots,x_n)$ $=$ $ f(x^{-1},\dots,x_n^{-1}).$ 

Then we have:
\begin{align}
& \chi D_1(q,t) \chi = t^{n-1} D_1(q^{-1},t^{-1}), \\ 
& \chi P_{\s} (q,t) = P_{\s^*}(q^{-1},t^{-1}), \qquad \s \in \LC_n^{(n)}. \label{eq:s*}
\end{align}
Here for $\s = (\s_1,\s_2,\dots,\s_n)$ we used the notation: $\s^* = (-\s_n,-\s_{n-1},\dots,-\s_1).$ From (\ref{eq:s*}) it follows that
\begin{equation}
\chi P_{\s}^{ (\gamma,N)} = P_{\s^*}^{ (\gamma,N)}. \label{eq:P*}
\end{equation}

\section{Dynamical Correlation Functions for $N=2$}
In this section we will apply results obtained thus far in order to compute Spin-Density and Density two-point Dynamical Correlation Functions in the fermionic Spin Calogero-Sutherland Model. So as to avoid cumbersome  technical details we will consider only the case where the spin has two values, i.e. $N=2$ and the number of particles in the Model $n$ is an even number such that $n/2$ is odd.     
\subsection{A parameterization of Energy, Momentum and Spin eigenvalues} \label{sec:parameterization}
Let $\{ X_{k}^{(\beta,2)} \; | \; k \in \LC_n^{(1)} \}$ be the $A(\gl_2;\beta)$-eigenbasis of the space $F_{2,n}$ constructed in section \ref{sec:a-eigenbasis}. 

The energy operator $\frac{2\pi^2}{L^2} H_{\beta,2}$ (\ref{eq:H}), the momentum operator $P = \frac{2\pi}{L} \sum_{i=1}^n z_i\p/\p z_i $ and the spin operator $S = \sum_{i=1}^n \s_i^z/2,$ where $\s_i^{z}$ are Pauli matrices,  all belong to the commutative family $A(\gl_2;\beta).$ Therefore their eigenvalues in the basis   $\{ X_{k}^{(\beta,2)} \; | \; k \in \LC_n^{(1)} \}$ can be read from the formulas (\ref{p4iii}) with the following result:
\begin{align}
& \frac{2\pi^2}{L^2} H_{\beta,2} X_{k}^{(\beta,2)} = \frac{2\pi^2}{L^2}E_{\beta,2}(k) X_{k}^{(\beta,2)}, \\
& \text{where} \quad E_{\beta,2}(k) = \sum_{i=1}^n \ov{k_i}^2 + \beta\sum_{i=1}^n(2i-n-1)\ov{k_i} + \frac{\beta^2n(n^2-1)}{12}; \label{eq:energyei}\\
& P X_{k}^{(\beta,2)} = P(k) X_{k}^{(\beta,2)}, \label{momentumei}\\ 
& \text{where} \quad P(k) = \frac{2\pi}{L}\sum_i^n \ov{k_i}; \\
& S X_{k}^{(\beta,2)} = S(k) X_{k}^{(\beta,2)}, \\ 
& \text{where} \quad S(k) = \frac{1}{2} \sum_{i=1}^n \delta(\un{k_i} = 1) - \delta(\un{k_i} = 2). \label{eq:spinei}
\end{align}
A ground state of the Model is defined as a state with the lowest energy eigenvalue among states with zero momentum. Under our assumption that $n$ is even and $n/2$ is odd, the ground state is unique and is found to be 
\begin{equation}
X_{k^0}^{(\beta,2)} \quad \text{with}\quad k^0 = (k^0_1,\dots,k^0_n) = (M,M-1,\dots,M-n+1) \quad \text{where} \; M = \frac{n}{2} + 1.
\end{equation}
From the expansion (\ref{p4ii}) we find that 
\begin{equation}
X_{k^0}^{(\beta,2)} = \vac(\frac{n}{2} + 1) = u_{k^0_1}\wedge u_{k^0_2} \wedge \cdots \wedge u_{k^0_n}.
\end{equation}
When the number of particles $n$ is even, any state $\vac(M)$ (\ref{eq:vac}) with even $M$ is a basis in a one-dimensional trivial representation of $Y(\gl_2;\beta).$  In particular, the spin of any such state is zero.    

For $k \in \LC_n^{(1)}$ define, as in sec. \ref{sec:basis}, a sequence $\s$ $\in$ $\LC_n^{(n)}$ by $ k_i = \s_i + k^0_i.$ Then the isomorphism $\Omega_{M-n+1}$ with  $ M = n/2 + 1$  maps $X_{\s + k^0}^{(\beta,2)}$ into $ P_{\s}^{(2\beta + 1,2)}.$ In particular, the ground state is mapped into 1 and a state $X_{\lambda + k^0}^{(\beta,2)}$ where $\lambda$ is a partition is mapped into the $\gl_2$-Jack Polynomial $  P_{\lambda}^{(2\beta + 1,2)}.$ For a state of this form we will find it convenient  to express the corresponding eigenvalues of energy, momentum and spin in terms of the diagram associated with the partition $\lambda.$ 
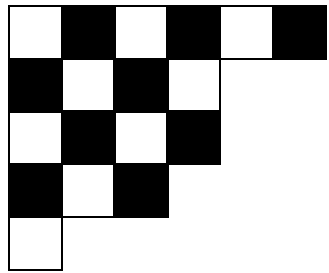
\begin{center}
\begin{picture}(120,100)
\put(0,100){\line(1,0){120}}
\put(0,80){\line(1,0){120}}
\put(0,60){\line(1,0){80}}
\put(0,40){\line(1,0){80}}
\put(0,20){\line(1,0){60}}
\put(0,0){\line(1,0){20}}
\put(0,0){\line(0,1){100}}
\put(20,0){\line(0,1){100}}
\put(40,100){\line(0,-1){80}}
\put(60,100){\line(0,-1){80}}
\put(80,100){\line(0,-1){60}}
\put(100,100){\line(0,-1){20}}
\put(120,100){\line(0,-1){20}}


\multiput(20,80)(40,0){3}{\rule{20pt}{20pt}}

\multiput(0,60)(40,0){2}{\rule{20pt}{20pt}}

\multiput(20,40)(40,0){2}{\rule{20pt}{20pt}}

\multiput(0,20)(40,0){2}{\rule{20pt}{20pt}}
\end{picture} \\

\end{center}
Let us color the diagram of $\lambda$ by white and black colors in the checkerboard order so that the square $(1,1)$ is colored white. An example of this coloring is given on the picture above which represents the partition $\lambda = (6,4,4,3,1).$ Then a square $s\in \lambda$ is white(black) if and only if the content of this square $c(s)$ is even(odd). Let $W_{\lambda}(B_{\lambda})$ be the subset of all white(black) squares in $\lambda$. In the notation of sec. \ref{sec:notations} we have $W_{\lambda} = C_2(\lambda).$ With this coloring we have for the eigenvalues of spin, momentum, and normalized energy ${\cal E}(k) = \frac{2\pi^2}{L^2}( E_{\beta,2}(k)- E_{\beta,2}(k^0)):$   
\begin{gather}
 S(\lambda + k^0) = |W_{\lambda}| - |B_{\lambda}|, \label{eq:spinei1}\\
 P(\lambda + k^0) = \frac{2\pi}{L} |W_{\lambda}|, \\ 
 {\cal E}(\lambda + k^0) = \frac{2\pi^2}{L^2}\left( n_w(\lambda') - (2\beta + 1) n_w(\lambda) + \left( \frac{n}{2}(2\beta + 1) + \frac{1}{2}\right)|W_{\lambda}|\right). 
\end{gather}
In the last formula $\lambda'$ is the  partition conjugated to $\lambda$ and $n_w(\lambda)$ which is analogous to $n(\lambda) = \sum (i  - 1)\lambda_i$ is defined as follows:  
\begin{equation}
n_w(\lambda) = \sum_{s \in W_{\lambda}} l'(s) = \sum (i  - 1)w_i(\lambda),
\end{equation}
where $w_i(\lambda)$ is the number of white squares in the $i$th row of $\lambda.$ Note that we have
\begin{equation}
n_w(\lambda') = \sum_{s \in W_{\lambda}} a'(s) = \sum_{i : \odd}w_i(\lambda)^2 - w_i(\lambda) + \sum_{i : \even}w_i(\lambda)^2.
\end{equation}
Finally, from the expressions (\ref{eq:energyei} - \ref{eq:spinei}) we find that if  $ \lambda^* = (-\lambda_n,-\lambda_{n-1},\dots,-\lambda_1)$ then  
\begin{eqnarray}
S(\lambda^* + k^0)& = -&S(\lambda + k^0), \label{eq:spinreflect} \\ 
P(\lambda^* + k^0)& = -&P(\lambda + k^0), \label{eq:momentumreflect} \\
{\cal E}(\lambda^* + k^0)& = &{\cal E}(\lambda + k^0). \label{eq:energyreflect}\end{eqnarray}

\subsection{Dynamical Correlation Functions} \label{sec:correlation}
We will consider two-point correlation functions of the following operators: 
\begin{eqnarray}
& \text{Spin-Density:} \quad & s(0,0) =  \sum_{i=1}^n \delta(y_i)\s_i^z/2 =  \frac{1}{L}\sum_{m\in \zint}  \sum_{i=1}^n  z_i^m  \s_i^z/2, \label{eq:sden} \\
& \text{Density:} \quad & \rho(0,0) = \sum_{i=1}^n \delta(y_i) - \frac{n}{L} = \frac{1}{L}\sum_{m=1}^{\infty} \sum_{i=1}^n ( z_i^m + z_i^{-m} ). \label{eq:rho}
\end{eqnarray}
Also, as an intermediate step in the computation  we will need to consider correlation functions of the  operators
\begin{equation}
\quad s_{\pm}(0,0) =  \sum_{i=1}^n \delta(y_i)\s_i^{\pm} =  \frac{1}{L}\sum_{m\in \zint}  \sum_{i=1}^n  z_i^m \s_i^{\pm}. \label{eq:spmden}
\end{equation}
Now let $s$ be an integer and define
\begin{gather}
 J_s^{+} = \sum_{i=1}^n z_i^{-s-1} \s_i^+, \quad J_s^{-} = \sum_{i=1}^n z_i^{-s} \s_i^-, \\  
 J_s = J_s^{+} + J_s^{-}, \qquad   K_s = \sum_{i=1}^n z_i^{-s}.
\end{gather}
So that $ s_{\pm}(0,0) = \frac{1}{L}\sum_{s\in \zint} J_s^{\pm} .$ Acting on a wedge in $F_{2,n}$ with operators $J_s$ and $K_s$ we have:
\begin{align}
& J_s u_{k_1}\wedge u_{k_2}\wedge \cdots \wedge u_{k_n} = \sum_{i=1}^n u_{k_1}\wedge u_{k_2}\wedge \cdots \wedge  u_{k_i + 2s+1}\wedge \cdots\wedge u_{k_n}, \label{eq:Ju}\\ & K_s u_{k_1}\wedge u_{k_2}\wedge \cdots \wedge u_{k_n} = \sum_{i=1}^n u_{k_1}\wedge u_{k_2}\wedge \cdots \wedge  u_{k_i + 2s}\wedge \cdots\wedge u_{k_n}. \label{eq:Ku}
\end{align}
Let $f$ and $g$ be eigenvectors of the spin operator in $F_{2,n}$ such that $ S f = s(f) f$ and $S g = s(g) g,$ where $s(f)$ and $s(g)$ are corresponding  eigenvalues. Since the operator $S$ is self-adjoint relative to the scalar product ${\sprod{\cdot}{\cdot}}_{\beta,2}$ we have: 
\begin{equation}
{\sprod{f}{J_s^{\pm} g }}_{\beta,2} = \delta\left( s(f) - s(g) = \pm 1\right) {\sprod{f}{J_s g }}_{\beta,2}.
\end{equation}
Now observe that (\ref{eq:Ju}, \ref{eq:Ku}) imply that the isomorphism $\Omega_{M-n+1}$ transforms the operators $J_s$ and $K_s$ into odd and even power-sums respectively:  
\begin{align}
&\Omega_{M-n+1}J_s\Omega_{M-n+1}^{-1} = \sum_{i=1}^n x_i^{2s + 1} = p_{2s + 1}, \\
& \Omega_{M-n+1}K_s\Omega_{M-n+1}^{-1} = \sum_{i=1}^n x_i^{2s} = p_{2s },
\end{align}
so that 
\begin{align}
& {\sprod{f}{J_s^{\pm} g }}_{\beta,2} = \delta\left( s(f) - s(g) = \pm 1\right)
{\sprodd{ \Omega_{M-n+1}(f)}{p_{2s + 1}\Omega_{M-n+1}(g) }}_{\beta,2}, \\
& {\sprod{f}{K_s g }}_{\beta,2} = \delta\left( s(f) - s(g) = 0\right){\sprodd{\Omega_{M-n+1}(f)}{p_{2s}\Omega_{M-n+1}(g) }}_{\beta,2}.
\end{align}
In virtue of these relations computation of matrix elements of the Spin-Density and Density operators is reduced to computation of matrix elements of power sums. Because of this from now on the logic of our computation follows exactly the logic of computation of the Density Correlation Function in the scalar Calogero-Sutherland Model \cite{Ha,Konno}.

Let us denote by $t$ and $x$ the time and the space coordinates in a two-point Correlation Function. Taking formulas discussed so far in this section into account we have: 
\begin{gather}
\langle s_{\mp}(x,t)s_{\pm}(0,0)\rangle \equiv  \\ 
\frac{  {\sprod{\vac(\frac{n}{2}+1)}{\exxp^{-i t\frac{2\pi^2}{L^2}H_{\beta,2}-ixP} s_{\mp}(0,0) \exxp^{i t\frac{2\pi^2}{L^2}H_{\beta,2}+ixP} s_{\pm}(0,0) \vac(\frac{n}{2}+1)}}_{\beta,2}}{{\sprod{\vac(\frac{n}{2}+1)}{\vac(\frac{n}{2}+1)}}_{\beta,2}} = \\
\sum_{\s \in \LC_n^{(1)}} \frac{ {\sprod{\vac(\frac{n}{2}+1)}{s_{\mp}(0,0) X_{\s + k^0}^{(\beta,2)}}}_{\beta,2} {\sprod{X_{\s + k^0}^{(\beta,2)}}{s_{\pm}(0,0)\vac(\frac{n}{2}+1)}}_{\beta,2}}{   {\sprod{\vac(\frac{n}{2}+1)}{\vac(\frac{n}{2}+1)}}_{\beta,2} {\sprod{X_{\s + k^0}^{(\beta,2)}}{X_{\s + k^0}^{(\beta,2)}}}_{\beta,2}} \exxp^{it{\cal E}(\s + k^0) + i x P(\s + k^0)} = \\ 
= \frac{1}{L^2}
\sum\begin{Sb}\s \in \LC_n^{(1)} \\ S(\s + k^0) = \pm 1 \end{Sb} 
\frac{ {\sprodd{ P_{\s}^{(2\beta + 1,2)}}{ \sum_{m : \odd} ( p_m + p_{-m} )}}_{\beta,2}^2 } { {\sprodd{ 1 }{ 1 }}_{\beta,2}{\sprodd{P_{\s}^{(2\beta + 1,2)}}{P_{\s}^{(2\beta + 1,2)}}}_{\beta,2} }\exxp^{it{\cal E}(\s + k^0) + i x P(\s + k^0)}.
\end{gather}
Now using  (\ref{eq:p=P+}), (\ref{eq:P*}) and (\ref{eq:spinreflect} - \ref{eq:energyreflect}) we obtain for the Spin-Density Correlation Function: 
\begin{gather}
\langle s(x,t) s(0,0)\rangle = \frac{1}{4}\langle s_{-}(x,t) s_{+}(0,0)\rangle + \frac{1}{4}\langle s_{+}(x,t) s_{-}(0,0)\rangle = \\
= \frac{1}{2 L^2}\sum\begin{Sb} \lambda,\: |\lambda| : \odd  \\ S(\lambda + k^0) = \pm 1 \end{Sb}
\left( \chi_{\lambda}^{(2\beta + 1,2)}\right)^2 \frac{{\sprodd{P_{\lambda}^{(2\beta + 1,2)}}{P_{\lambda}^{(2\beta + 1,2)}}}_{\beta,2}}{{\sprodd{ 1 }{ 1 }}_{\beta,2}}\exxp^{it{\cal E}(\lambda + k^0)}\cos(x P(\lambda + k^0)).\nonumber
\end{gather}
In view of  (\ref{eq:norm}), (\ref{eq:psexp3},  \ref{eq:psexp4}) and  (\ref{eq:spinei1}) the last expression may be written as follows   
\begin{gather}
\langle s(x,t) s(0,0)\rangle =  \label{eq:ss}\\
= \frac{1}{2 L^2}\sum\begin{Sb} \lambda,\: |\lambda| : \odd  \\ |W_{\lambda}| - |B_{\lambda}| = \pm 1 \\  |W_{\lambda}| = |H_2(\lambda)| + 1 \end{Sb}
\frac{ {\displaystyle \prod_{ s \in W_{\lambda}\setminus(1,1)}} c(s;2\beta + 1)^2 }{ {\displaystyle\prod_{ s \in H_2(\lambda)}} h_{\lambda}^*(s; 2\beta + 1) h^{\lambda}_*(s; 2\beta + 1) }Z_{\lambda}(\beta,n)\;\exxp^{it{\cal E}(\lambda + k^0)}\cos(x P(\lambda + k^0)), \nonumber\\ 
\text{where} \qquad Z_{\lambda}(\beta,n) = \prod_{s \in W_{\lambda}}\frac{ a'(s) + (2\beta + 1)(n - l'(s))}
{ a'(s) + 1 + (2\beta + 1)(n - l'(s)-1)}.\nonumber
\end{gather}
In complete analogy with the Spin-Density Correlation Function, we find for the Density Correlation Function:
\begin{gather}
\langle \rho(x,t) \rho(0,0)\rangle =  \\
=\frac{2}{ L^2}\sum\begin{Sb} \lambda,\: |\lambda| : \even  \\ S(\lambda + k^0) = 0 \end{Sb}
\left( \chi_{\lambda}^{(2\beta + 1,2)}\right)^2 \frac{{\sprodd{P_{\lambda}^{(2\beta + 1,2)}}{P_{\lambda}^{(2\beta + 1,2)}}}_{\beta,2}}{{\sprodd{ 1 }{ 1 }}_{\beta,2}}\exxp^{it{\cal E}(\lambda + k^0)}\cos(x P(\lambda + k^0)).\nonumber
\end{gather}
Which in view of (\ref{eq:norm}), (\ref{eq:psexp1} , \ref{eq:psexp2}) , (\ref{eq:spinei}) we may write  as   
\begin{gather}
\langle \rho(x,t) \rho(0,0)\rangle =  \label{eq:dd}\\
= \frac{2}{ L^2}\sum\begin{Sb} \lambda,\: |\lambda| : \even  \\ |W_{\lambda}| - |B_{\lambda}| = 0 \\  |W_{\lambda}| = |H_2(\lambda)| \end{Sb}
|\lambda |^2 \frac{  {\displaystyle \prod_{ s \in W_{\lambda}\setminus(1,1)}} c(s;2\beta + 1)^2 }{ {\displaystyle\prod_{ s \in H_2(\lambda)}} h_{\lambda}^*(s; 2\beta + 1) h^{\lambda}_*(s; 2\beta + 1) }Z_{\lambda}(\beta,n)\;\exxp^{it{\cal E}(\lambda + k^0)}\cos(x P(\lambda + k^0)), \nonumber\\ 
\text{where} \qquad Z_{\lambda}(\beta,n) = \prod_{s \in W_{\lambda}}\frac{ a'(s) + (2\beta + 1)(n - l'(s))}
{ a'(s) + 1 + (2\beta + 1)(n - l'(s)-1)}. \nonumber 
\end{gather}
Notice that according to the definition (\ref{eq:rcontent}) the summands in formulas for  Correlation Functions vanish if the diagram of $\lambda$ contains the ( white for integer $\beta$ ) square $( 2 , 2\beta + 2 ).$ This observation must facilitate a computation of the thermodynamic limit. In this paper we do not consider this problem.  

\section{Discussion}
In this paper we have constructed an isomorphism between the space of states of the $\glN$-invariant Calogero-Sutherland Model and the space of Symmetric Laurent Polynomials. With this isomorphism we have shown that  the $\glN$-invariant Calogero-Sutherland Models with $N=1,2,3,\dots\;$ when considred in an appropriate  unified  framework, which is the framework of Symmetric Polynomials,  manifest themselves as limiting cases of the same entity, which is the commuting family of Macdonald Operators. Macdonald Operators depend on two parameters $q$ and $t$. And the Hamiltonian of $\glN$-invariant Calogero-Sutherland Model belongs to a degeneration of this family in the limit when both $q$ an $t$ approach the $N$th elementary root of unity:
\begin{equation}
q = \o_N p, \quad  t = \o_N p^{ N \beta + 1 }, \quad p \rightarrow 1, \nonumber
\end{equation}
where $\o_N = \exp(\frac{2 \pi i }{N} )$ and $\beta$ is the coupling constant of the $\glN$-invariant Calogero-Sutherland Model. This picture provides  a generalization of the well-known situation in the case of Scalar  Calogero-Sutherland Model $(N=1).$ 

In the limit the commuting family of Macdonald Operators is identified with the maximal commutative subalgebra $A(\glN)$ in the action of the Yangian algebra  $Y(\glN)$ on the space of states of the  $\glN$-invariant Calogero-Sutherland Model. The limits of Macdonald Polynomials which we call for obvious reasons $\glN$-Jack Polynomials are eigenvectors of this subalgebra and, by definition, form  Yangian Gelfand-Zetlin bases in irreducible components of the Yangian action. 

In view of the isomorphism between the space of states of the $\glN$-invariant Calogero-Sutherland Model and the space of Symmetric Laurent Polynomials the latter admits an action of the Yangian $Y(\glN).$ This fact is not surprising for it is known that the level-1 Fock space module of the affine Lie algebra $\widehat{\gl}_N$ admits a Yangian action coming from that of the Yangian action of the Calogero-Sutherland Model in certain projective limit $n \rightarrow \infty$ ( see \cite{U} for the $\widehat{\gl}_2$ case ) and the Fock space module is isomorphic by the Fermion-Boson correspondence to the ring of Symmetric Functions \cite{KR}. In fact the isomorphism of our present paper is to  be regarded as nothing more than a finite-size version of the  Fermion-Boson correspondence.

The connection with the Macdonald Operators seems to be more mysterious. It provides another example of a situation where degree of a root unity appears as rank of a Lie algebra. To understand this phenomenon better it would be desirable to understand of which algebra the Yangian is the limit when $q$ and $t$ approach the root of unity.       

The $\glN$-Jack Polynomials describe the orthogonal eigenbasis of  $\glN$-invariant Calogero-Sutherland Model in exactly the same way as Jack Polynomials describe the orthogonal eigenbasis of the Scalar Model $(N=1).$ For each known property of Macdonald Polynomials there is a corresponding property of $\glN$-Jack Polynomials. As a simplest application of these properties we compute two-point Dynamical Spin-Density and Density Correlation Functions in the $\gl_2$-invariant Calogero-Sutherland Model at integer values of the coupling constant. 

In this paper we did not consider the problem of computation of the Green's Functions. However we expect that the Cauchy formula for the $\glN$-Jack Polynomials:
\begin{gather}      
 \sum_{\lambda} b_{\lambda}^{(\gamma ,N)}  P_{\lambda}^{(\gamma ,N)} ( x_1,\dots,x_n)  P_{\lambda}^{(\gamma ,N)}( y_1,\dots,y_n) = \prod_{ i,j =1}^n ( 1 - x_i^N y_j^N)^{\beta}  ( 1 - x_i y_j) \nonumber  \\
 \text{ where $\gamma = N\beta + 1$ and (see \ref{eq:b}) } \qquad   b_{\lambda}^{(\gamma,N)} = \prod_{s \in \lambda}  b_{\lambda}^{(\gamma,N)}(s)  \nonumber 
\end{gather}
and its dual 
\begin{equation}
\sum_{\lambda} P_{\lambda}^{(\gamma,N)}(x_1,\dots,x_n)P_{\lambda'}^{(\gamma^{-1},N)}(y_1,\dots,y_n) = \prod_{i,j =1}^n ( 1 + x_i y_j)  \nonumber 
\end{equation}
 will play a role in a solution of this problem together with an appropriate evaluation formula for  $\glN$-Jack Polynomials.

\mbox{} \\

\appendix

\noindent {\bf {\Large Appendix }}
\section{Proof of the statements (\ref{p2i} ) - (\ref{p2v})}
The statement (\ref{p2i}) follows from the well-known construction of Yangian representations by the fusion procedure ( see e.g. \cite{Jimbo}). So we will not discuss proof of this statement here. 

To prove the rest of the statements, for any two increasing sequences ${\bold a}= (a_1,a_2,\dots,a_m)$, $ {\bold b} = (b_1,b_2,\dots,b_m)$ where $ m,a_i,b_i \in \setN$ define (Cf.\cite{NT}):  
\begin{equation}
Q_{{\bold a},{\bold b}}(u) := \sum_{w\in S_m}(-1)^{l(w)} T_{a_1,b_{w(1)}}(u)
T_{a_2,b_{w(2)}}(u-1)\cdots T_{a_m,b_{w(m)}}(u-m+1).
\end{equation}
With this definition:
\begin{equation}
A_m(u) = Q_{{\bold m},{\bold m}}(u), \qquad \text{where}\quad {\bold m}=\{1,2,\dots,m\}, \qquad (m=1,2,\dots,N). \label{eq:A=Q}
\end{equation}
According to Proposition 1.11 in \cite{NT}:
\begin{equation}
\Delta^{(n)}(A_m(u)) = \sum_{{\bold a}^{(1)},\dots, {\bold a}^{(n-1)}}
Q_{{\bold m},{\bold a}^{(1)}}(u)\otimes Q_{{\bold a}^{(1)},{\bold a}^{(2)}}(u)\otimes \cdots \otimes Q_{{\bold a}^{(n-1)},{\bold m}}(u), \label{eq:A-coprod}
\end{equation}
where the summation is taken over all increasing sequences of integers $1,2,\dots, N $ of length $m$. 

By considering the action of the operator $ E_{1,1}+ 2E_{2,2}+\dots+N E_{N,N} $ $\in$ $End(V)$ we find for $g\in \real$, $a\in \setN$ that:   
\begin{equation}
\pi(g)(Q_{{\bold a},{\bold b}}(u))v_a \in \real[[u^{-1}]] v_{a+|{\bold b}|-|{\bold a}|}, 
\end{equation}
and hence that for any $a \in \setN^n$  the expression 
\begin{equation}
\left(\pi(f_1)\otimes\pi(f_2)\otimes \cdots \otimes \pi(f_n)\right) \Delta^{(n)}( A_m(u) ) v(a)  
\end{equation}
is a linear combination of $ v(b) $ such that the $b \in  \setN^n$ has as its elements  
\begin{equation}
b_i = a_i + |{\bold a}^{(i)}|-|{\bold a}^{(i-1)}|, \qquad (i=1,2,\dots,N),  
\end{equation}
for  certain ${\bold a}^{(1)},\dots, {\bold a}^{(n-1)}$ appearing in (\ref{eq:A-coprod}) and ${\bold a}^{(0)} ={\bold a}^{(n)} = {\bold m}.$ Since for any such $b$ we have 
\begin{equation}
b_1+b_2+\cdots+b_i = a_1+a_2+\cdots+a_i + |{\bold a}^{(i)}| - |{\bold m}|, \qquad (i=1,2,\dots,n), 
\end{equation} 
and since in the last equation $ |{\bold a}^{(i)}| - |{\bold m}| > 0 $ unless ${\bold a}^{(i)}= {\bold m}$, we find by taking (\ref{eq:A=Q}) into account that\begin{align}
& \left(\pi(f_1)\otimes\pi(f_2)\otimes \cdots \otimes \pi(f_n)\right) \Delta^{(n)}( A_m(u) ) v(a)  =  \\ & = \left(\pi(f_1)(A_m(u))\otimes\pi(f_2)(A_m(u))\otimes \cdots \otimes \pi(f_n)(A_m(u))\right) v(a) + \sum_{b > a} c(u;f;a,b) v(b)  = \nonumber \\ & = \sum_{b \geq a} c(u;f;a,b) v(b), \nonumber 
\end{align}
where $c(u;f;a,b) \in \real[[u^{-1}]]$ and 
\begin{equation}
c(u;f;a,a) = \prod_{i=1}^n\frac{u+f_i +\delta(a_i\leq m)}{u+f_i} =  A_m(u;f;a).
\end{equation}

Now by the definition (\ref{eq:phi-def}) we have 
\begin{equation}
\varphi(a) = v(a) + \sum_{b>a}e(a,b) v(b), \qquad ( a \in T_p, \quad e(a,b) \in \{0,\pm 1\} ), \label{eq:phi=v>}
\end{equation}
and hence: 
\begin{multline}
\left(\pi(f_1)\otimes\pi(f_2)\otimes \cdots \otimes \pi(f_n)\right) \Delta^{(n)}( A_m(u) ) \varphi(a) = \\ = c(u;f;a,a) v(a) + \sum_{b > a} d(u;f;a,b) v(b), \qquad ( d(u;f;a,b) \in \real[[u^{-1}]]). \label{eq:Aphi1}
\end{multline}

On the other hand  the subspace $(\otimes^n V)_p$ is left invariant by the Yangian action, so that 
\begin{multline}
\left(\pi(f_1)\otimes\pi(f_2)\otimes \cdots \otimes \pi(f_n)\right) \Delta^{(n)}( A_m(u) ) \varphi(a) = \\ =  \sum_{b} g(u;f;a,b) \varphi(b), \qquad ( g(u;f;a,b) \in \real[[u^{-1}]]).
\end{multline}
Comparing this equation with (\ref{eq:Aphi1}) and using (\ref{eq:phi=v>}) we get
\begin{multline}
\left(\pi(f_1)\otimes\pi(f_2)\otimes \cdots \otimes \pi(f_n)\right) \Delta^{(n)}( A_m(u) ) \varphi(a) = \\ =  A_m(u;f;a)\varphi(a)+ \sum_{b>a} c(u;f;a,b) \varphi(b), \qquad ( c(u;f;a,b) \in \real[[u^{-1}]]). \label{eq:A-triang}
\end{multline}

Now define the sequence of real numbers $ h^{(1)}, h^{(2)},\dots, h^{(M)}$ by
\begin{equation}
h^{(s)} := f_{q_{s-1}+1}, \quad (  s=1,2,\dots,M ).
\end{equation} 
And define for any  $a \in T_p$:
\begin{equation}
l_m^{(s)} = \sum_{i=q_{s-1}+1}^{q_s} \delta(a_i \leq m ), \qquad ( m=1,2,\dots,N; \; s=1,2,\dots,N). \label{eq:lseq}
\end{equation}
Then for the eigenvalue  $A_m(u;f;a)$ we have 
\begin{equation}
 A_m(u;f;a)= \prod_{s=1}^M \frac{u+1+h^{(s)}}{u+1+h^{(s)}-l_m^{(s)}}.
\end{equation}
The correspondence 
\begin{equation}
 T_p \; \rightarrow \; \{ l_m^{(s)} \; | \; s=1,2,\dots,M;\quad m=1,2,\dots,N \} 
\end{equation}
given by  the relation  (\ref{eq:lseq}) is bijective. The conditions (\ref{eq:fbar1}, \ref{eq:fbar2}) and the inequalities $ 0\leq l_m^{(s)} \leq q_s - q_{s-1} $ imply that 
\begin{equation}
h^{(s)}-l^{(s)}_m-h^{(s+1)}+l^{(s+1)}_m > 0, \qquad (s=1,2,\dots,M-1; \quad m=1,2,\dots,N),    
\end{equation}
so that the $N$-tuple of rational functions in the variable $u$: $A_1(u;f;a),A_2(u;f;a),\dots,A_N(u;f;a)$ determines the set  $\{ l_m^{(s)} \; | \; s=1,2,\dots,M,\quad m=1,2,\dots,N \} $ uniquely. This proves the statement (\ref{p2v}). Now the  statements (\ref{p2ii}),(\ref{p2iii}) and (\ref{p2iv}) follow from (\ref{eq:A-triang}) and (\ref{p2v}). Thus the proof is finished.

\section{Proof of the equality (\ref{eq:Phi=Psi})}
Using the expression (\ref{eq:Uop}) for the operator $U(s;\beta)$ we have
\begin{equation}
 \Phi^{(\beta)}(s,a) = \sum_{t \sim s} E_t^{(\beta)}(z)\otimes R_t^{(\beta)}\varphi(a),
\end{equation}
where $t \sim s$ means that $t$ belongs to the set of all distinct rearrangements of the sequence $s \in \LC_n^{(N)}.$
In view of the triangularity (\ref{eq:nsJtri}) of the Non-symmetric Jack Polynomial  $E_t^{(\beta)}(z)$ we may split the $E_t^{(\beta)}(z)$ with $t \sim s$ as follows:
\begin{gather}
E_t^{(\beta)}(z) = E_t^{(\beta)}(z)' + E_t^{(\beta)}(z)'', \\
\text{where} \quad  E_t^{(\beta)}(z)' = \sum_{ r \preceq t, \; r \sim s} e^{(\beta)}_{tr} z^r , \quad \text{ and } \quad  E_t^{(\beta)}(z)'' = \sum_{ m \in \LC_n^{(N)},\; m < s,\; r \sim m} e^{(\beta)}_{tr} z^r. 
\end{gather}
Both of the vectors
\begin{equation}
\sum_{t \sim s} E_t^{(\beta)}(z)'\otimes R_t^{(\beta)}\varphi(a) \quad \text{and}\quad  \sum_{t \sim s} E_t^{(\beta)}(z)''\otimes R_t^{(\beta)}\varphi(a)
\end{equation}
belong to the subspace $F_{N,n}$ since $\Phi^{(\beta)}(s,a)$ belongs to  $F_{N,n}$ and monomials $z^r$ which appear in the decomposition of $E_t^{(\beta)}(z)'$ as a vector in $\cz\otimes(\otimes^n V)$ are distinct from monomials $z^r$ which appear in the decomposition of $E_t^{(\beta)}(z)''.$  

Taking into account that $ R_s^{(\beta)} = 1$ we may write:
\begin{equation}
\sum_{t \sim s} E_t^{(\beta)}(z)'\otimes R_t^{(\beta)}\varphi(a) = z^s \otimes \varphi(a) + \sum_{t \sim s,\; t \prec s} z^t \otimes \bar{\varphi}_t(a), \quad ( \bar{\varphi}_t(a) \in \otimes^n V).
\end{equation}
Since the expression above is a vector in $F_{N,n}$ and by using the definition of the vector $\varphi(a)$ in (\ref{eq:phi-def}) we obtain 
\begin{equation}
\sum_{t \sim s} E_t^{(\beta)}(z)'\otimes R_t^{(\beta)}\varphi(a) = A_n( z^s \otimes v(a) )  = (-1)^{\frac{n(n-1)}{2}} \hat{u}_k  \label{eq:e1prime}
\end{equation}
where $A_n$ (\ref{eq:antisymm}) is the operator of total antisymmetrization in the space $\cz\otimes (\otimes^n V),$ and as in the paragraph preceding the equation (\ref{eq:Phi=Psi}) we have  $k = (k_1,k_2,\dots,k_n) \in \LC_n^{(1)}$ such that 
\begin{equation}
 s  =  (\ov{k_n},\ov{k_{n-1}},\dots,\ov{k_1}), \quad a  = (\un{k_n},\un{k_{n-1}},\dots,\un{k_1}).
\end{equation}
Furthermore, the expansion of  the vector 
\begin{equation}
\sum_{t \sim s} E_t^{(\beta)}(z)''\otimes R_t^{(\beta)}\varphi(a) \label{eq:e2prime}
\end{equation}
in $\cz\otimes(\otimes^n V)$ contains only monomials $z^r$ such that $ r^{+} < s.$ Therefore expansion of (\ref{eq:e2prime}) in the basis $\{ \hat{u}_l \: | \: l \in \LC_n^{(1)} \}$ contains only  normally ordered wedges $\hat{u}_l$ such that $ \ov{l} > \ov{k} .$ Taking this, and (\ref{eq:e1prime}) into account we have
\begin{equation}
(-1)^{\frac{n(n-1)}{2}}\Phi^{(\beta)}(s,a) = \hat{u}_k  + \sum_{l \in \LC_n^{(1)},\; \ov{l} > \ov{k}} \varphi_{kl}^{(\beta)} \hat{u}_l  \qquad ( \varphi_{kl}^{(\beta)} \in \real).
\end{equation}
However according to (\ref{p1i}) an eigenvector of the Hamiltonian $H_{\beta,N}$  with the above expansion in the basis $\{ \hat{u}_l \: | \: l \in \LC_n^{(1)} \}$  is unique and equals to $\Psi_k^{(\beta)}.$ This proves (\ref{eq:Phi=Psi}).

\end{document}